\documentclass[twocolumn,aps,prd,longbibliography]{revtex4-1}
\usepackage{amsmath,latexsym}
\usepackage{xcolor}
\usepackage{tgbonum}
\usepackage{graphicx}
\usepackage{dcolumn}
\usepackage{bm}
\usepackage{times} 
\usepackage{capt-of}
\usepackage{booktabs}
\usepackage{varwidth}
\usepackage{graphics}
\usepackage{graphicx}
\usepackage{array}
\usepackage{dcolumn}
\usepackage{bm}
\usepackage{float}
\usepackage{enumitem}
\usepackage[colorinlistoftodos, textwidth=4cm, shadow]{todonotes}
\usepackage{amssymb}

\begin{document}
\title{Borderline Magnetism: How Does Adding Magnesium to Paramagnetic CeCo$_3$ Make a 450 K Ferromagnet with Large Magnetic Anisotropy?}
\author{Tribhuwan Pandey}
\email{pandeyt@ornl.gov}
\affiliation{Material Science and Technology Division, Oak Ridge National Laboratory Oak Ridge 37831 TN USA}
\author{David S. Parker}
\email{parkerds@ornl.gov}
\affiliation{Material Science and Technology Division, Oak Ridge National Laboratory Oak Ridge 37831 TN USA}
\date{\today}

\begin{abstract}
A recent experimental study (Phys. Rev. Appl. 9, 024023, 2018) on paramagnetic CeCo$_3$ finds that Magnesium alloying induces a ferromagnetic transition with intrinsic properties large enough for permanent magnet applications. Here we explain these surprising results \textit{via} a first principles study of the electronic structure and magnetism of Magnesium-alloyed CeCo$_3$. We find the origin of this Magnesium-induced ferromagnetic transition to be Stoner physics - the substantial increase in the Fermi-level density-of-states $N(E_F)$ with Mg alloying. Our calculations suggest that both Ce and Co atoms are important for generating large magnetic anisotropy suggesting the viability of Co-3$d$, and Ce-4$f$ interaction for the generation of magnetic anisotropy in magnetic materials. These results offer a new route to the discovery of ferromagnetic materials and provide fundamental insight into the magnetic properties of these alloys.
\end{abstract}

\maketitle
\section{Introduction}
Magnetism and magnetic materials have been known colloquially since antiquity, with lodestone$^\prime$s magnetic properties, such as its attraction to Iron, noted in ancient texts \cite{mattis1965theory,coey2010magnetism}. More recently, with the industrial, scientific and technological revolutions occurring since the latter half of the 19$^{\mathrm{th}}$ century has come increasing interest in {\it explaining} magnetic behavior. This has often been related to the properties of {\it exchange}: the process by which the Pauli exclusion principle, combined with local crystal structure, results in the moments of magnetic ions such as Iron or Manganese aligning parallel, or antiparallel, to their nearest neighbor magnetic ions. Non-collinear states and more complex ordering patterns are also frequently observed with neutron scattering \cite{shull1951neutron,lovesey1984theory,shull1995early,may2017magnetic,may2016competing}.  

Density functional theory (DFT), as originally formulated by Kohn and Sham \cite{hohenberg1964inhomogeneous,kohn1965self,lecture1999electronic,kohn1996density} and implemented by Perdew \cite{perdew1992atoms,perdew2005prescription,perdew1996generalized} and others \cite{kresse1996efficient,singh2006planewaves}, has become a useful tool for the study of such systems, leading to the ability to understand and even {\it predict} the behavior of magnetic materials. This has led to increasing interest in ``high-throughput" calculations \cite{drebov2013ab,korner2016theoretical,sanvito2017accelerated,curtarolo2012aflowlib}, in which a large number of materials are rapidly screened \textit{via} DFT in an attempt to find properties technologically useful for applications such as permanent magnets. In general, these calculations usually predict the proper (or occurrence of a) magnetic ground state and often attain good agreement ($i.e.$ within 10 percent) with experimentally measured magnetic moment values, while calculations of magnetic anisotropy are more difficult, particularly for rare-earth compounds such as the permanent magnet workhorse Nd$_2$Fe$_{14}$B \cite{coey1986intrinsic,susner20172flux,tatetsu2016first,yi2017multiscale}.

Recently Canfield $et.al.$~\cite{tej-CeCo3-Mg} found ferromagnetic behavior in the Cerium-Cobalt intermetallic CeCo$_3$ upon alloying with Mg. Perhaps surprisingly, despite containing 75 atomic percent of the ferromagnetic Co, which orders at 1388 K~\cite{heller1967experimental,keffer1966handbuch} CeCo$_3$ does not order magnetically. However, alloying of this compound with Mg not only renders it ferromagnetic, but does so with a Curie point T$_{C}$ as high as 450 K and a large 50 K magnetic anisotropy of 2.2 MJ/m$^3$.   These Curie point and anisotropy values are in the range of potential permanent magnets, although this specific material would likely require substantial  further optimization for actual usage as a permanent magnet.  How does adding a non-magnetic element to paramagnetic CeCo$_3$ yield a 450 K ferromagnet with large magnetic anisotropy?

Here, we answer this basic question. We find its resolution in Stoner physics - the substantial increase in the Fermi-level density-of-states $N(E_F)$ with Mg alloying - and in the magnetic anisotropy of the Ce $4f$ and Co 3$d$ orbitals. This study is organized as follows. First we describe the calculations methods, the essential input parameters and approximations adopted to achieve the desired numerical convergence. Next, we discuss the calculated properties of the base compound CeCo$_3$, finding that the properties of CeCo$_3$ are sensitive to the exchange-correlation potential (Local Spin Density Approximation (LDA) or Generalized Gradient Approximation (GGA)). We then describe the magnetic properties of the Mg substituted compound (Ce$_2$Co$_9$Mg). We show that the experimentally observed enhancement in magnetic properties are described by the Stoner picture. The experimental results on the Mg-substituted alloys are fully corroborated by our first principles calculations, and we show that a similar enhancement in magnetic properties arises from alloying with Ca. Finally we show that these and related results~\cite{shtender2017crystal} on Mg-alloyed NdCo$_3$ and other ferromagnets open a new pathway for the development of high performance magnets. 

\begin{figure}[!h]
\includegraphics[width=0.9\columnwidth]{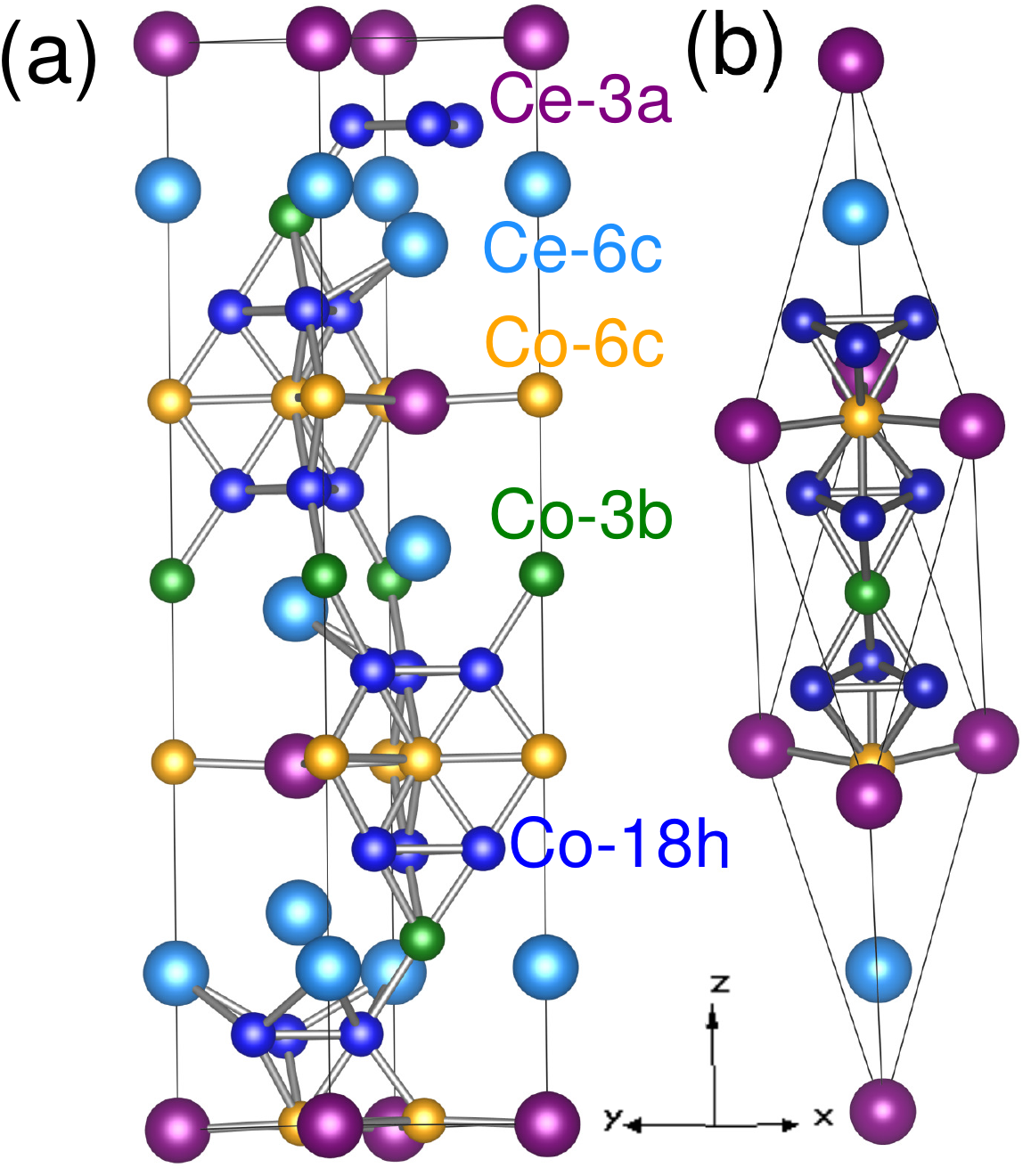}
\caption{Schematic representation of the crystal structures of CeCo$_3$. Nonequivalent Ce atoms are shown in magenta (Ce-3$a$) and cyan color (Ce-6$c$). The Nonequivalent Co-3$b$, and 6$c$, and 18$h$ sites are shown by orange, green, and blue spheres, respectively. (b) The primitive cell used for calculating the properties.}
\label{fig1}
\end{figure}

\section{Theoretical Methods}
 The calculations were performed by using the all electron density functional code WIEN2K~\cite{wien2k} at the experimental lattice parameters~\cite{tej-CeCo3-Mg}, which are listed in Table~\ref{table1}. The internal coordinates were relaxed until forces on all the atoms were less than 1 mRyd/Bohr. Muffin tin sphere radii of 2.5 for Ce, 2.2 for Co, and 2.1 for Mg were used. For good basis set convergence, a RK$_{max}$ value of 7.0 was used. The magnetic anisotropy energy (MAE) is obtained by calculating the total energies of the system with spin orbit coupling (SOC) as K$_1$ = E$_{a}$ $-$ E$_{c}$, where E$_{a}$ and E$_{c}$ are the total energies for the magnetization oriented along the in-plane [1$\bar{1}$0] and out-of-plane[111] directions. For the structure relaxation 1000 \textbf{k}-points were used in the full Brillouin zone. MAE as a small quantity (of the order of meV), can sensitively depend on the number of \textbf{k}-points used. The difference in MAE calculated using 6000 and 4000 \textbf{k}-points was less than 4 \% and all the MAE results reported here use 4000 \textbf{k}-points. The convergence of MAE with respect to \textbf{k}-points is discussed in the Appendix. The accurate calculation of MAE for rare earth transition metal complexes within the conventional LDA/GGA framework is challenging. Within these DFT approaches the partially filled rare earth $f$ states are pinned at the Fermi level, which results in incorrect properties. This problem can be remedied by treating the $f$ electrons as unhybridized core states (the open core approximation~\cite{brooks1991origin}).   However this approximation often results in the wrong value of the MAE~\cite{larson2003calculation,larson2003calculation1}. Another approach is to consider $f$ electrons as valence electrons and introduce Hubbard U correction to the \textit{f} orbitals which splits the \textit{f} bands into lower and upper Hubbard bands.  This is known as the LDA$+$U approach and we follow this approach here. Here, LDA$+$U corrections were included for the Ce 4$f$ orbitals using the self interaction correction (SIC)\cite{anisimov1993density,liechtenstein1995density,shick1999implementation,madsen2005charge} for the double-counting correction. In all the calculations Hund$^\prime$s coupling parameter J was set to 0 eV. As the Co states in Ce$_2$Co$_9$Mg are not localized, a U correction was not used for Co sites. As discussed in the following sections, introducing a U for Ce $f$ orbitals is crucial in the prediction of the correct magnetic anisotropy and orbital moments. We find that MAE calculated with U = 1.5 eV is in good agreement with experiments.

\section{Results and Discussions}

\subsection{Properties of CeCo$_3$} 
The unit cell and primitive cell (used in our simulations) are shown in Figures~\ref{fig1} (a) and (b). CeCo$_3$ crystallizes in a rhombohedral structure with space group $R$-$3m$. In this structure Co atom has three independent sites namely 3$b$, 6$c$, and 18$h$, whereas Ce atom has two independent site (3$a$, and 6$c$). As shown in Figure~\ref{fig1} both Ce sites has different arrangement of nearest neighbors. While Ce-3$a$ site is surrounded by six Co-6$c$ (first nearest) and twelve Co-18$h$ (second nearest) neighbors, the Ce-6$c$ site has six first nearest (Co-18$h$) and only three (Co-3$b$) second nearest neighbors. Numerous prior studies~\cite{mazin2004density, khmelevskyi2005magnetism, aguayo2004n,larson2004magnetism,sieberer2006magnetic} on the weak itinerant magnetic systems have shown that the choice of functional can be very important in predicting the correct magnetic ground state. Hence, first a comparative analysis of LDA and GGA functional for CeCo$_3$ was performed. These results are discussed in the Appendix Figure~\ref{fig7}. While both LDA and GGA calculations favor a ferromagnetic state, at experimental lattice parameters within LDA, the energy difference between ferromagnetic and non-magnetic state ($E_{\mathrm{FM}}-E_{\mathrm{NM}}$) is only $-$1.9 meV on per Co basis. On the other hand within GGA this difference is about $-$21 meV per Co atom. This result is consistent with the well-known tendency of the GGA to exhibit stronger magnetic instabilities then the LDA. Also the Co magnetic moment calculated at experimental lattice parameters within LDA (0.33 $\mu_B$) and GGA (0.99 $\mu_B$) functionals are very different. Experimentally CeCo$_3$ is often characterized as Pauli paramagnetic~\cite{buschow1980magnetic,tej-CeCo3-Mg}, although there is older literature~\cite{lemaire1966magnetic} finding evidence for magnetic character in CeCo$_{3}$.  This suggests that CeCo$_3$ is at or near a magnetic instability, which is better represented in this case by the LDA approach, with its much smaller magnetic energy ($-$1.9 meV/Co). Unless otherwise stated, all the calculations presented in the following sections are therefore performed using the LDA at the experimental lattice constants from Ref.~\cite{tej-CeCo3-Mg}.

\begin{table}[!ht]
\begin{center}
\caption{The experimental lattice constants from Ref.~\cite{tej-CeCo3-Mg} employed in our calculations along with the calculated spin (within the sphere~\cite{mm-note}) ($\mathrm{\mu_{Co/Ce}^{spin}}$) and orbital magnetic moments ($\mathrm{\mu_{Co/Ce}^{orb}}$), and magnetic anisotropy (K$_1$). The experimentally measured magnetic moments (m$_{tot}^{exp}$) and magnetic anisotropy (K$_1$ (exp.)) are also shown for comparison.The calculations for CeCo$_3$ are done within LDA, and calculations for Ce$_2$Co$_9$Mg are done within LDA$+$SOC$+$U ( U$_{\mathrm{Ce}}$ = 1.5 eV). K$_1^{OC}$ denotes MAE value calculated within open-core approximation, where Ce 4$f$ electrons are treated as core electrons. }
\label{table1}
\begin{tabular}{c| c| c}
\hline
Compound  & CeCo$_3$ & Ce$_2$Co$_9$Mg \\
\hline 
$a$ (\AA) & 4.94 & 4.92 \\
$c$ (\AA)& 24.64 & 24.01\\
$c/a$ & 4.98 & 4.87  \\
Volume (\AA$^3$)&521.33 & 504.75\\
$\mathrm{\mu_{Co}^{spin} (\mu_B}$) & 0.33& 1.17\\
$\mathrm{\mu_{Co}^{orb} (\mu_B}$) &  & 0.09\\
$\mathrm{\mu_{Ce}^{spin} (\mu_B}$) &$-$0.20 & $-$0.51\\
$\mathrm{\mu_{Ce}^{orb} (\mu_B}$) & & 0.23 \\
$\mathrm{m_{tot} (\mu_B}$/per f.u.)& 0.71 &10.04 \\
m$_{tot}^{exp}(\mu_B$/per f.u.)&0 $\lesssim$ 0.40~\cite{buschow1980magnetic,tej-CeCo3-Mg} & 8.0~\cite{tej-CeCo3-Mg} \\
K$_1$ (MJ/m$^3$)  & & 2.10 \\
K$_1^{OC}$ (MJ/m$^3$) & & 0.22 \\
K$_1$ (exp.) (MJ/m$^3$) & & 2.20~\cite{tej-CeCo3-Mg} \\
\hline
\end{tabular}
\end{center}
\end{table}
\subsection{Effect of Mg substitution on magnetic properties of CeCo$_3$}
\begin{figure}[ht!]
\centering
\includegraphics[width=0.9\columnwidth]{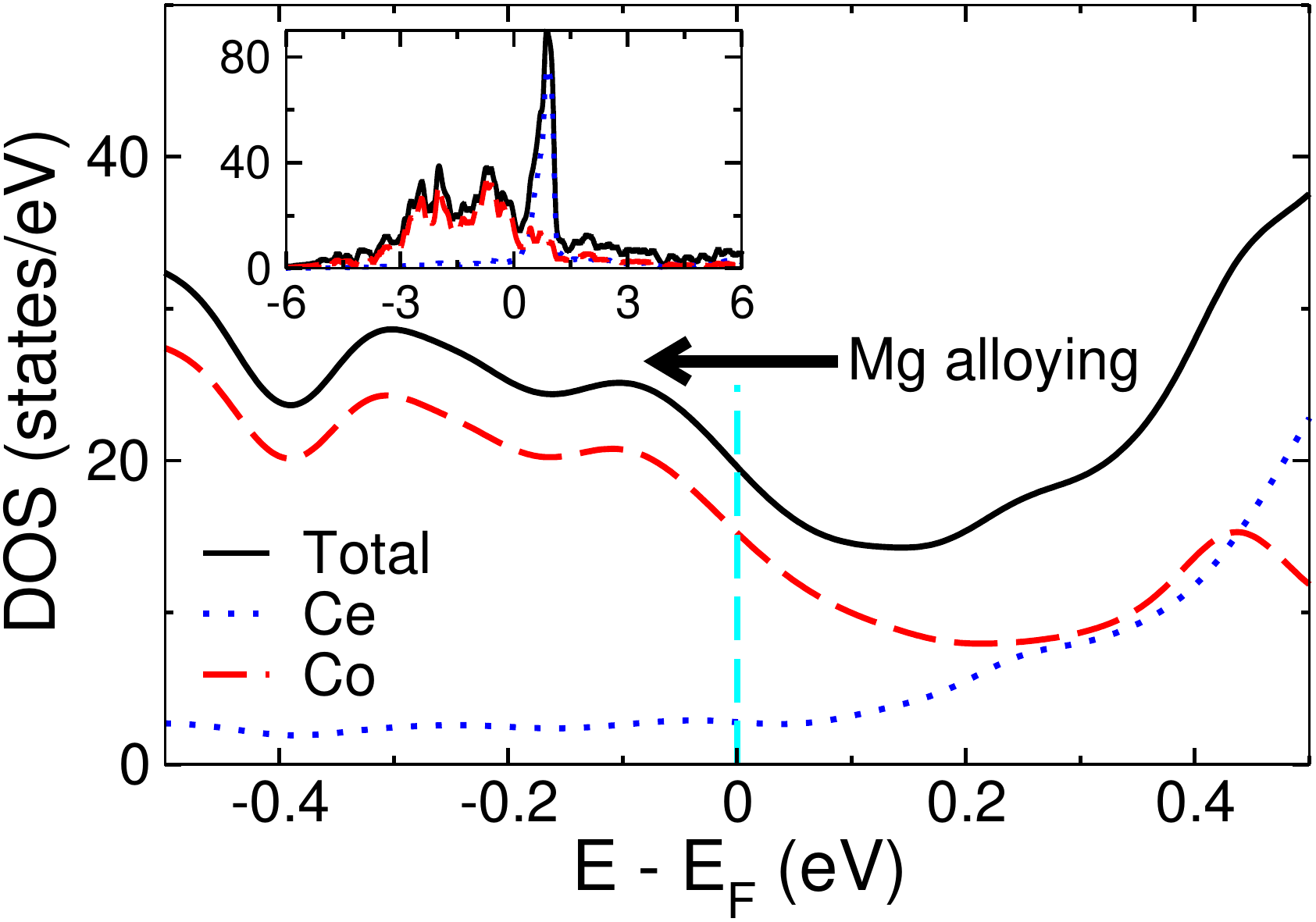}
\caption{The calculated non-magnetic density of states of CeCo$_{3}$ with LDA functional. For simplicity neither spin-orbit coupling nor a Hubbard U are included. Note the rapid increase in both total and Co DOS as the energy is reduced below E$_{F}$ (vertical cyan line), as occurs with alloying of Mg.}
\label{fig2}
\end{figure}

As mentioned previously, the primary question that theory should answer is how Mg alloying transforms the Pauli paramagnet CeCo$_3$ into a ferromagnet with properties - $T_c$ of 450 K and a low-temperature anisotropy field of 10 T - comparable to those of known permanent magnets. The basic reason for this behavior can in fact be seen from Figure~\ref{fig2}. There we plot the calculated non-magnetic density-of-states (DOS) of the base compound CeCo$_3$, calculated within the LDA approximation. For simplicity, in this calculation we omit spin-orbit coupling and the Hubbard U which will play an important role in detailed comparison with experiment later.

The plot depicts the calculated DOS in a narrow window around the Fermi level (E$_{F}$) (main plot) and for several eV around E$_{F}$ in the inset. From the inset we see two main features - the large Ce peak (in blue), from the unoccupied localized $4f$ states, centered around 1 eV above E$_{F}$, and the Co DOS, mainly from $-$4 eV to $+$1 eV (in red).  The Co DOS provides the majority of the spectral weight around E$_{F}$, suggesting the Co as being the ``driving force" for magnetism in the Ce$_{3-x}$Mg$_x$Co$_9$ alloy system.
In the main plot we see that for energies below the E$_{F}$, the DOS increases rapidly, both for the total and for the Co. Indeed, at an energy 0.18 eV below E$_{F}$, the values for these quantities are nearly 1.5 times higher than the Fermi-level values of 19.0 and 15.3 per eV-unit cell (note that the primitive cell contains 9 Co, or 3 formula units). This immediately suggests that ``hole-doping" (borrowing a term from semiconductor physics) will tend to increase the Fermi-level DOS. Now, the Stoner criterion~\cite{stoner1939collective} states that a ferromagnetic instability occurs when the condition $IN(E_{F}) > 1$ is fulfilled. Here $I$ is the Stoner parameter, which we compute by calculating average exchange splitting ($\Delta E_{ex}$) and by using the relation $\Delta E_{ex} = I\mathrm{m}_{avg}$~\cite{ortenzi2011competition}. Here m$_{avg}$ is the average magnetic moment of Co atom in CeCo$_3$. We find \textit{I} as 0.56 eV, which is quite close to 0.49 eV value for elemental Co~\cite{janak1977uniform}. Additional details involving the calculation of \textit{I} are listed in the Appendix. For our purposes $N(E_F)$ should be understood as the total Co DOS, on a per Co basis. Hence hole doping should be understood as likely to render CeCo$_{3}$ more magnetic.

Our calculations find the valency of Ce in CeCo$_3$ to be dependent on the approximation used (see Figure\ref{fig8} and corresponding discussion in the Appendix).  This may originate in valence fluctuations, although a detailed discussion of this issue is beyond the scope of this paper. Such valence fluctuations do occur in rare-earth compounds~\cite{SalesPhD,coey1993ce2fe17,jarlborg2014role,alam2014mixed,matsumoto2009first,sales1976demagnetization,sales1975susceptibility,sales1977model}. In our conventional LDA (without SOC and without U) calculations Ce in CeCo$_3$ is tetravalent, which is consistent with the experimentally observed non-magnetic behavior, while Mg is nearly universally in a $+$2 charge state. Therefore, Mg donates 2 {\it less} electrons to the bands than Ce, and thus tends to result in effective ``hole doping", an increase in Fermi-level DOS, and thereby an increase in ferromagnetic behavior, consistent with experiment. For CeCo$_3$ the total Co-non spin-polarized N($E_{F}$) is 15.7 states/eV-unit cell or 1.7 states/eV on per Co atom basis, indicating that the Stoner criterion (taking \textit{I} as $\sim$ 0.56 eV as described in the Appendix) is very nearly satisfied ($IN\mathrm{(}E_{F}$) $\sim$ 0.98). On the other at 0.18 eV below the Fermi level the Stoner criterion ($IN\mathrm{(}E_{F}$) $\sim$ 1.47) is clearly fulfilled.

The previous analysis is done within a ``rigid-band" picture, in which the primary effect of Mg alloying is a reduction in the Fermi energy. However, one should not thereby argue that correlations are unimportant in this system - we will see that they are indeed important in generating magnetic anisotropy. However, the correlations here generally apply to the Ce atom, not to the Co atom. As in many rare-earth magnets, it is the $3d$ (in this case Co) atoms which are generally driving the magnetic {\it instability}. 
\subsection{Magnetic properties of Ce$_2$Co$_9$Mg}

\begin{figure}[ht!]
\centering
\includegraphics[width=0.85\columnwidth]{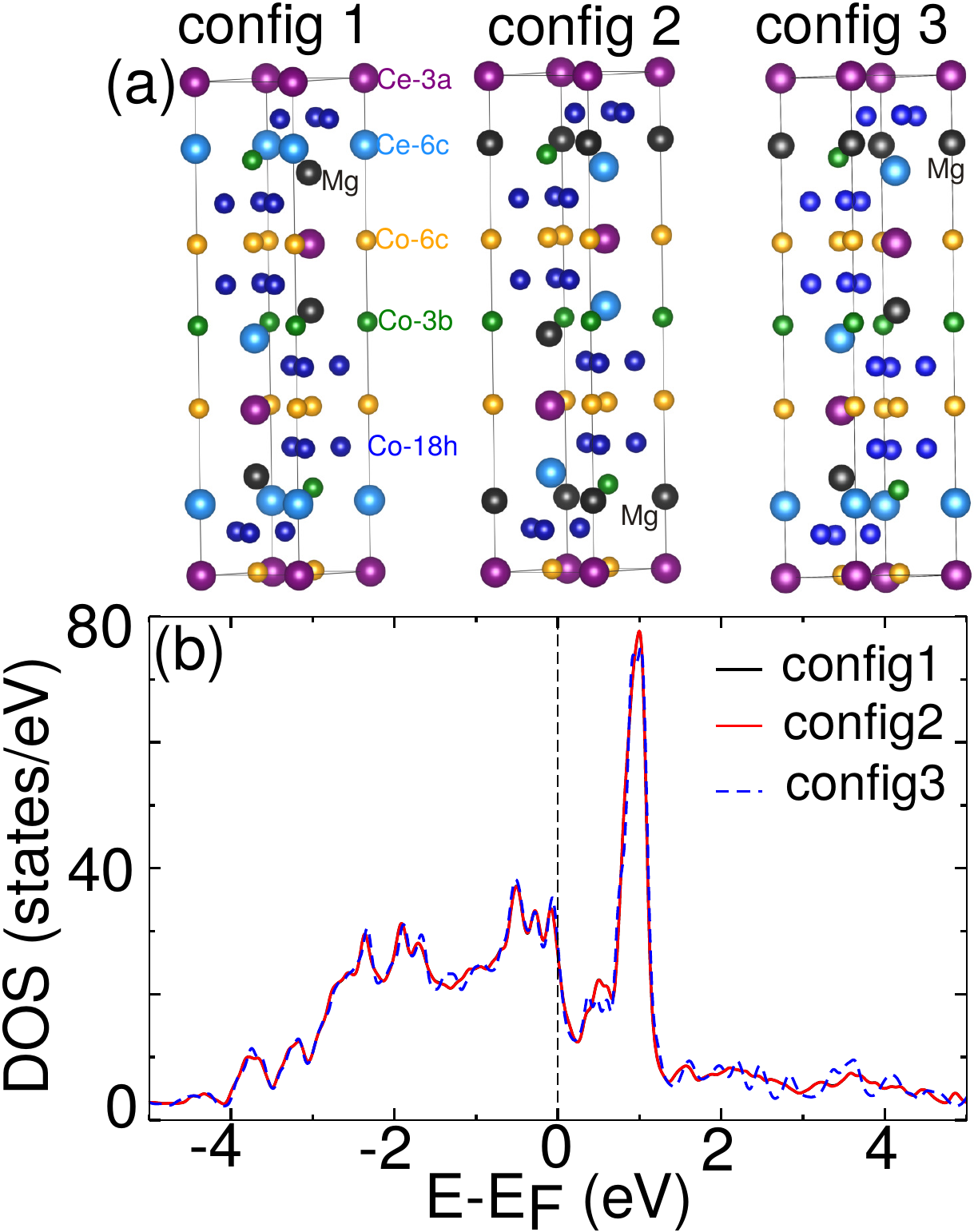}
\caption{(a) The configurations used for substituting Mg (shown by black spheres) in CeCo$_3$ unit cell. (b) The non spin polarized density of states calculated within LDA.}
\label{fig3}
\end{figure}
As discussed in the previous section based on electronic structure and Stoner theory analysis, the observed ferromagnetism in CeCo$_3$ upon Mg substitution can be explained based on Stoner physics. Next we calculate magnetic properties of Ce$_2$Co$_9$Mg. We begin with identifying the proper ground state structure for Ce$_2$Co$_9$Mg. The recent experimental efforts~\cite{tej-CeCo3-Mg} conclude that Mg alloying in CeCo$_3$ prefers to occupy Ce-6$c$ site and structure maintains its rhombohedral symmetry. To mimic such a structure Mg atoms were substituted in 50-50 fashion among six available Ce-6$c$ sites in the 36 atom unit cell. These configurations are presented in Figure~\ref{fig3} (a). All three structures were relaxed to their ground state by minimizing the total forces. While for configuration 1 and configuration 2 the rhombohedral symmetry is broken and structure transform into a less symmetry trigonal cell (with space group \textit{P3m1}), the configuration 3 maintains its rhombohedral symmetry (space group \textit{R-3m}). Energetically we find configuration 3 to be most stable,  whereas configuration 1 and 2 are higher in energy (unstable) by only 8.3 meV relative to configuration 3. To make sure that the calculated magnetic properties are independent of the configuration used, we next compare their density of states (DOS) near the Fermi level, which is shown in Figure~\ref{fig3} (b).  We find that the DOS of all three configurations are nearly identical. As the magnetic properties and MAE are dependent on the behavior of DOS in the vicinity of Fermi level, the nearly similar DOS implies that the calculated magnetic proprieties should be independent of configuration used. Also, as the experimental study finds the structure to be rhombohedral, configuration 3 was used in all the calculations for Ce$_2$Co$_9$Mg.

\begin{figure}[ht!]
\centering
\includegraphics[width=\columnwidth]{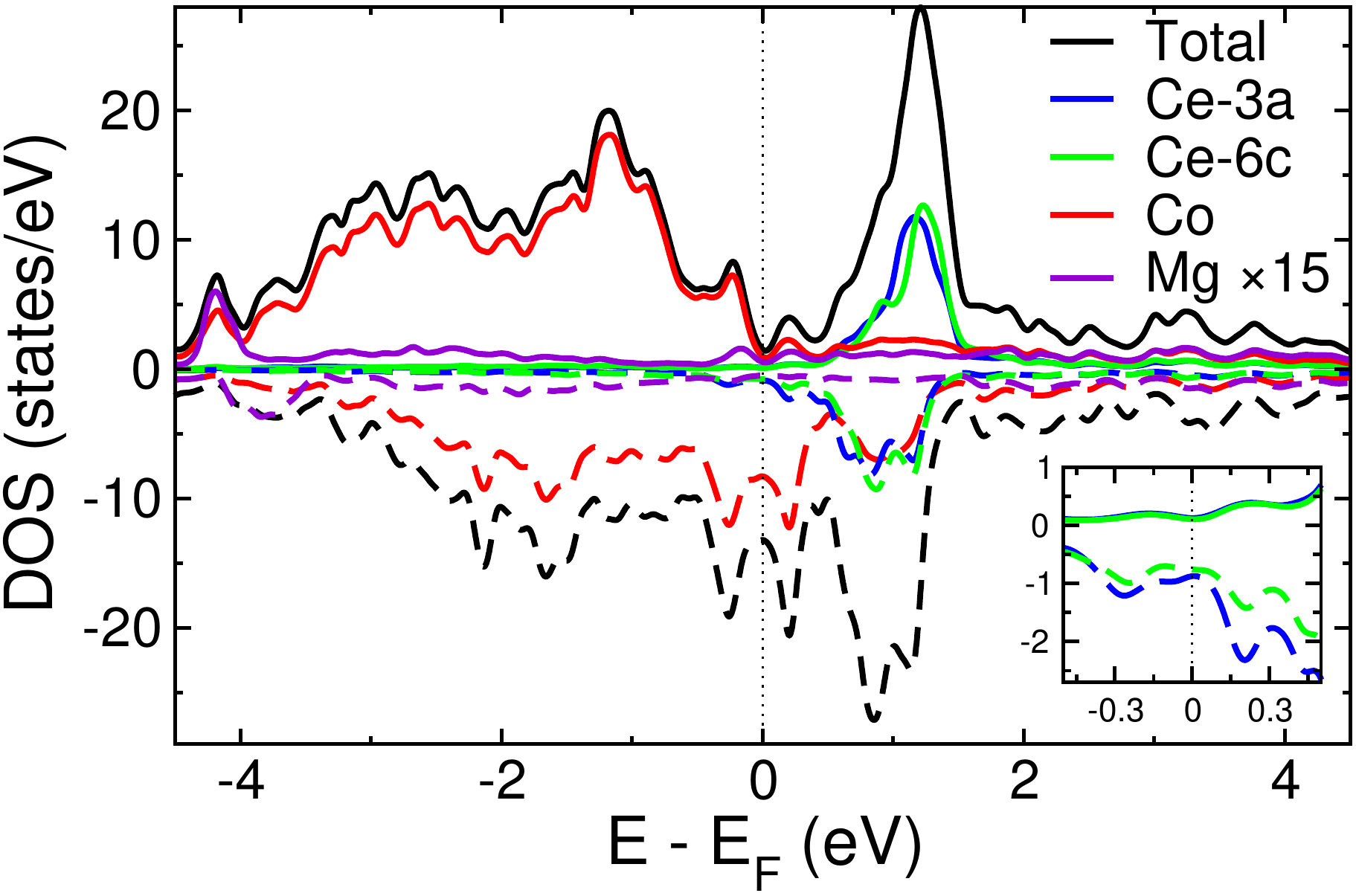}
\caption{The calculated density of states of Ce$_2$MgCo$_9$. The solid and dotted lines denote the density of states for spin up and spin down channels respectively. The Mg DOS is enlarged by 15 times for clarity. The Co DOS is averaged over all 9 Co atoms. Inset shows the enlarged Ce DOS near the Fermi level. The DOS is computed within LDA$+$SOC$+$U with U = 1.5 eV at Ce site.}
\label{fig4}
\end{figure}

In order to predict the accurate magnetic ground state of Ce$_2$Co$_9$Mg various collinear spin configurations with different spin arrangement of Co atoms were considered.  Nearly all of the configurations (with the exception of the ferrimagnetic state shown in Table~\ref{table2}) converged to the stable ferromagnetic state. As observed in the experimental measurements we find a ferromagnetic behavior in Ce$_2$MgCo$_9$, with a calculated average magnetic moment of 1.18 $\mu_B$ per Co as shown in Table~\ref{table1}. This includes an average orbital moment of $\sim$ 0.09 $\mu_B$ on the Co site. As can be seen from Table~\ref{table1} compared to base compound CeCo$_3$, upon Mg alloying (Ce$_2$Co$_9$Mg) the magnetic moment on Co site is enhanced by more than 3.5 times. Due to different site symmetry the Ce sites has different orbital moments of 0.20 $\mu_B$ for Ce-3$a$ and 0.16 $\mu_B$ for Ce-6$c$ sites. Upon introducing a Hubbard U the Ce orbital moment increases and for U = 1.5 it becomes 0.27 $\mu_B$ for Ce-3$a$ and 0.19 $\mu_B$ for Ce-6$c$ sites. U dependence of Ce orbital moment is discussed in the next section. As expected Ce atoms prefers to be anti-aligned with respect to Co with an average magnetic moment (including average orbital moment of 0.23 $\mu_B$) of $-$0.28 $\mu_B$ per Ce atom. This is similar to numerous other rare-earth magnet results~\cite{pandey_2_17_3_paper, pandey2018magnetic}. The calculated total magnetic moment of 10.04 $\mu_B$ per formula unit is in fair agreement with the 50 K experimental value of $\sim$ 8.0 $\mu_B$. For Ce$_{1.67}$Mg$_{1.33}$Co$_{9}$, this deviation in the predicted magnetic moment can be explained with the decreasing trend of saturation moment of Ce$_{3-x}$Mg$_{x}$Co$_{9}$ samples higher than  $x$ = 1.11 as discussed in Ref.~\cite{tej-CeCo3-Mg}.  We present the density of states (DOS) in Figure~\ref{fig4}. The magnetic moments on Co sites are somewhat lower than that observed in SmCo$_5$ and YCo$_5$~\cite{larson2003magnetic,zhu2014lda} (1.51 $\mu_B$ per Co). This is also corroborated by the DOS plot where the spin up DOS is substantially lower than the spin down DOS around E$_{F}$, reducing the magnetic moment. The enhanced ($\times$ 15) Mg DOS is also shown in Figure~\ref{fig4} (b). Although contribution on Mg DOS at Fermi level is small, there is some hybridization present with the neighboring Co and Ce atoms particularly in the majority spin channel. The Mg-$s$ states hybridize strongly with neighboring Co atoms at around $-$4 eV below the Fermi level. Both Ce sites has large DOS above the Fermi level. The enlarged Ce DOS in the vicinity of Fermi level is shown in the inset of Figure~\ref{fig4}. In the minority spin channel the Ce-3$a$ site has relatively larger DOS, which is consistent with the higher calculated magnetic moment. As explained before, alloying with Mg results in enhancement of Co-DOS in the vicinity of Fermi level, improving the magnetic properties as well the Co moments.

\begin{figure}[ht!]
\centering
\includegraphics[width=\columnwidth]{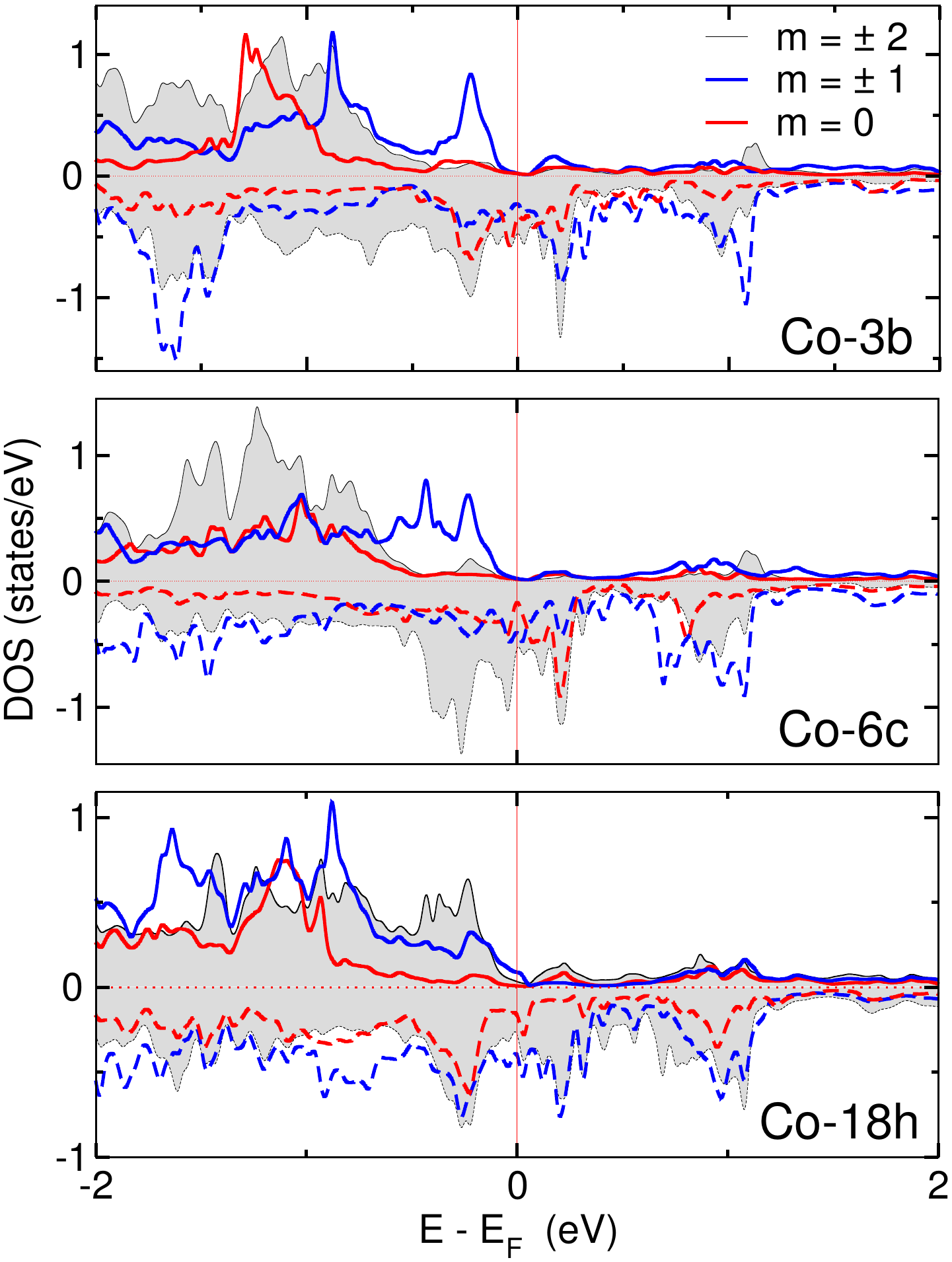}
\caption{The spin polarized partial density of states projected on the 3d states of non equivalent Co sites in Ce$_2$Co$_9$Mg. The black (gray filled), blue, and red colors indicate the contribution from $m = 2$ (degenerate d$_{xy}$ and d$_{x^2 - y^2}$ states), $m = 1$ ( degenerate d$_{xz}$ and d$_{yz}$), and $m = 0$ (d$_{z^2}$) states respectively. Positive and negative DOS represent majority and minority spin channels respectively. These calculations are done within LDA, without spin-orbit coupling and without U.}
\label{fig5}
\end{figure}

We examine the spin and orbital character of the orbitals near the Fermi level~\cite{daalderop1994magnetic,antropov2014constituents,ke2015band,ke2016origin}.  It is well-known that the magnetocrystalline anisotropy arises under the influence of spin-orbit coupling, and may be expressed as a second-order perturbative interaction between occupied and unoccupied states. Also as shown by Schilfgaarde $et. al.$~\cite{ke2015band},  {\it intra} $m$ transitions between the same (different) spin channel favor easy axis (planar) anisotropy, and {\it inter} $m$ transitions between the same (different) spin channel favor planar (easy-axis) anisotropy. The scalar-relativistic partial densities of states (PDOS) projected on the various non equivalent Co sites are shown in Figure ~\ref{fig5}. Without SOC five Co-$d$ orbitals split into three groups $m$ = 0 ($d_z^2$) , $m$ = $\pm$1 ($d_{yz}|d_{xz}$), and  $m$ = $\pm$2 ($d_{x^2-y^2}|d_{xy}$) and they are shown in Figure~\ref{fig5} by red, blue, and black (gray filled) lines respectively. In this plot we use the LDA, without spin-orbit or a Ce U.  Notably, all the $d$ orbitals hybridize with the neighboring atoms and overall, the DOS exhibits broad bandwidth (atomic like narrower bands are not present). As expected for all three non equivalent Co sites has large DOS in the majority spin channel just below the Fermi level, but very small DOS at the Fermi level itself. In particular for Co-3$b$ and Co-6$c$ sites contribution from $d_{xz}|d_{yz}$ orbitals ($m$ = $\pm$1) dominates, while for Co-18$h$  significant contribution from $d_{x^2-y^2}|d_{xy}$ orbitals is also present. The minority spin channel show significant DOS at and around the Fermi level for all three Co sites. Highest contribution from $d_{x^2-y^2}|d_{xy}$ along with d$_{z^2}$ for Co-3$b$, from $d_{x^2-y^2}|d_{xy}$ and d$_{xz}|d_{yz}$ for Co-6$c$, and from d$_{xz}|d_{yz}$ and $d_{x^2-y^2}|d_{xy}$ for Co-18$h$ can be seen. Overall a complex behavior is observed, making a quantitative analysis of the origin of the MAE difficult.  As described later, this difficulty is compounded by the substantial dependence of the calculated MAE on the value of U assumed for Ce.  What we can say is that multiple Co orbitals contribute to the magnetic behavior, likely including the MAE, in a manner specific to each distinct site, with the magnetic properties of the Co atoms substantially impacted by a large spin-down DOS near E$_{F}$.

\subsection{Magnetic anisotropy energy for Ce$_2$Co$_9$Mg}
\begin{figure}[ht!]
\centering
\includegraphics[width=\columnwidth]{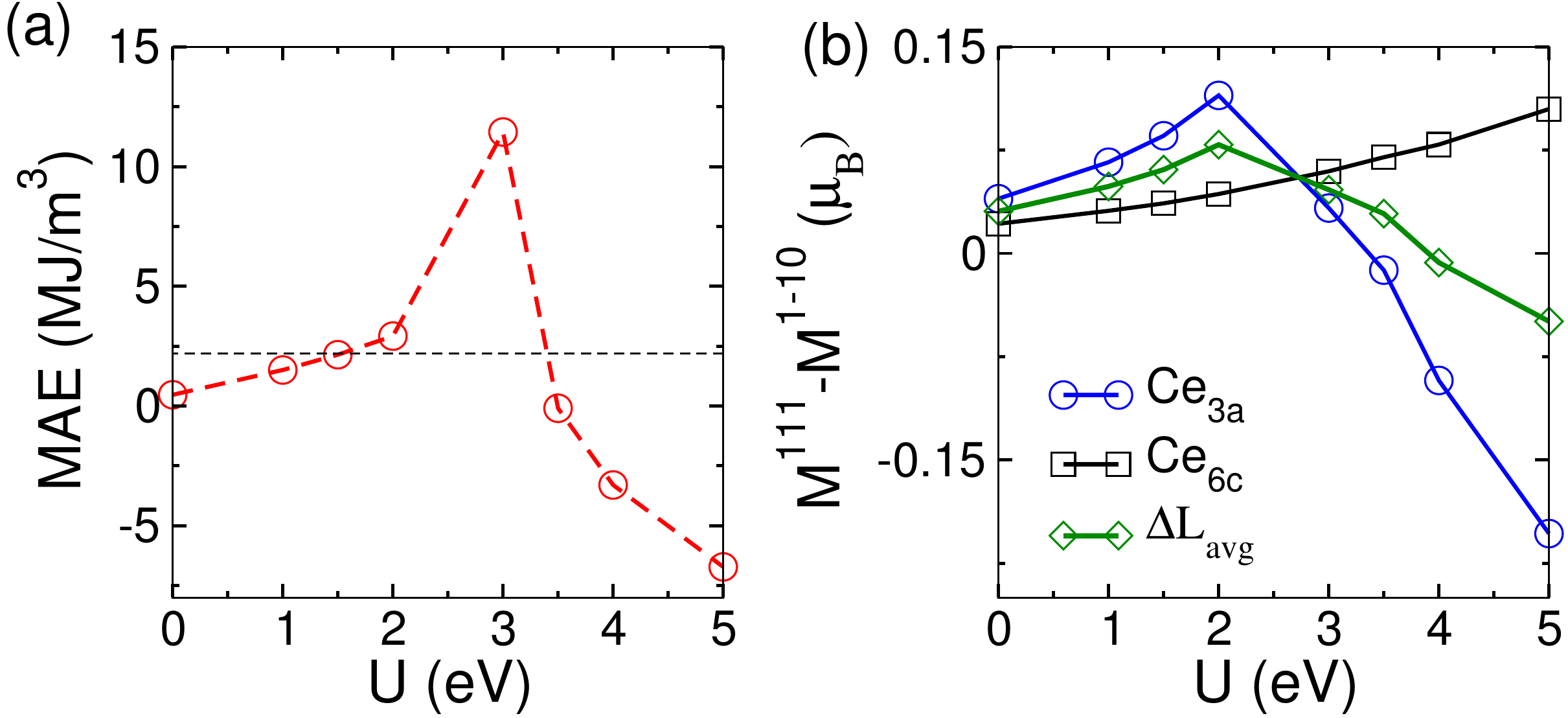}
\caption{(a) Dependence of MAE with respect to U parameter at Ce site. The circles denote the calculated data points. Note that the line is a guide to the eye. The experimental value of MAE~\cite{tej-CeCo3-Mg} is shown by dashed black line. (b) Variation of orbital moment anisotropy ($\Delta L$) at Ce-3$a$ and Ce-6$c$ site as a function of U at Ce site. Variation in average orbital moments ($\Delta L_{avg}$) is shown by diamonds. These calculations are done within LDA$+$SOC$+$U.}
\label{fig6}
\end{figure}

Next we calculate MAE for Ce$_2$Co$_9$Mg, which are tabulated in Table~\ref{table1}. First a calculation without including a Hubbard U parameter was performed. MAE calculated without including U at Ce site is about 0.46 MJ/m$^3$ much smaller than the experimental K$_1$ value of 2.2 MJ/m$^3$.
This suggests that to improve the agreement between theory and experiments a correlated approach is necessary. The dependence of MAE on the U-parameter is shown in Figure~\ref{fig6} (a). We find from LDA$+$U$+$SOC calculations that the MAE is sensitive to the U parameter used, and increases to 2.10 MJ/m$^3$ (2.25 meV/f.u) at U = 1.5 eV. This is in good agreement with the experimental value of 2.2 MJ/m$^3$. This is a somewhat smaller U value than typically taken for Ce, and for more typically employed values of U (3 or 4 eV) one finds very different values of the MAE, with the MAE even becoming negative for U past 3.5 eV. The strong dependence of MAE on U argues for the relevance of correlations in treating Ce here, and for the importance of Ce in generating magnetic anisotropy. To further show the importance Ce-4$f$ states for generating MAE,  we also calculated MAE by treating Ce-$f$ states as core electrons (known as open-core approximation). As shown in Table~\ref{table1} the MAE calculated within open-core approximation is very small (0.22 MJ/m$^3$) implying that Ce-$f$ valence electrons play an important role in magnetic properties of Ce$_2$Co$_9$Mg. The MAE can also be correlated to the anisotropy of orbital moments by following the Bruno$^\prime$s relation~\cite{bruno1989tight} MAE = $\sum_{j}\lambda_{j} \Delta L_{j} \cdot S_{j} $. Here $\lambda_{j}$, $\Delta L_{j}$, and $S_{j}$ refer to the spin orbital coupling constant, anisotropy of orbital moments, and spin moment of atomic species $j$. 

The dependence of orbital momentum anisotropy ($\Delta L)$ on U value for non-equivalent Ce sites is shown in Figure~\ref{fig6} (b). $\Delta L $ is computed by taking by taking the difference between orbital moments along out-of-plane [111] direction and in-plane [1$\bar{1}$0] direction. We find that Ce-3$a$ and Ce-6$c$ site exhibit very different orbital moment dependence on U.  At U = 0 both sites have nearly the same $\Delta L$. While for a U value in between 0 and 2 eV, the Ce-3$a$ site exhibits higher orbital moment along out-of-plane direction and  $\Delta L$ is positive, it becomes negative as U becomes larger than 2 eV. The situation is rather different for Ce-6$c$ site, and for all the U values investigated here $\Delta L$ remains positive and exhibits relatively much weaker dependence on U. This different orbital moment for different Ce sites suggest that the role of Ce in generating MAE for this compound is rather complex and site-specific.  The U dependence of average Ce $\Delta L_{avg}$ is also shown in Figure~\ref{fig6} (b). Due to its large $\Delta L$ the trend of $\Delta L_{avg}$ is governed by Ce-3$a$ site. For the most part the MAE follows the Bruno$^\prime$s relation and can be (qualitatively) explained by the trend in of $\Delta L$, with the only exception to this at U = 3 eV.

While the strength of SOC for 4$f$ rare earths electrons is approximately 0.5 eV, for 3$d$ transition metals it is an order of magnitude, or more smaller. This together with our calculated $\Delta L_{\mathrm{Ce}}$ = 0.060 $\mu_B$ (at U = 1.5 eV) and $\Delta L_{\mathrm{Co}}$ = 0.009 $\mu_B$, indicates that the majority of MAE should originate from Ce atoms. Usually the individual atomic contribution to the MAE can also be quantified by selectively switching off the SOC on the different atomic sites. However for Ce$_2$Co$_9$Mg such an analysis does not give consistent results and the total MAE calculated by adding the contribution from individual atomic sites is 60\% larger than the actual MAE (obtained by applying SOC at both Ce and Co sites simultaneously). This difference suggests that the analyzing the atomic origin of the MAE in this compound is far more complicated and considering cross Ce-Co spin orbit coupling terms may be essential. Nonetheless, while given their great spectral weight near $E_F$ in Figure~\ref{fig2} it is the Co atoms that drive the magnetic instability, both Co and Ce atoms may well be important for the MAE.
\subsection{Effect of Mg alloying on Curie point}
Perhaps the most remarkable observation of the recent experimental measurements~\cite{tej-CeCo3-Mg} is the considerable increase in Curie temperature of CeCo$_3$ by alloying with Mg. Generally, for a local-moment ferromagnetic system, the Curie temperature may be estimated by the energy difference ($\Delta$E = E$_{\mathrm{AF}}$ $-$ E$_{\mathrm{FM}}$) between an antiferromagnetic structure (AF) where the nearest-neighbors of all Co atoms are anti-aligned and the ferromagnetic (FM) ground state. While the rhombohedral structure leads to a magnetic frustration that prohibits such an arrangement, we have constructed a ferrimagnetic (FI) configuration that achieves the same basic purpose, and is described in the inset of Table~\ref{table2} for Ce$_2$Co$_9$Mg. In the mean-field local moment approximation the Curie temperature can then be estimated as one third of this energy difference, measured on a per Co basis. To quantify the effect of Mg on Curie point, in addition to CeCo$_3$ and Ce$_2$MgCo$_9$, we also study CeMg$_2$Co$_9$ (obtained by replacing a Ce by Mg in Ce$_2$MgCo$_9$). The resulting energy difference between FI structure and FM structure is listed in Table~\ref{table2}. In these calculations we include spin-orbit coupling and a Hubbard U. For CeCo$_3$ we could not stabilize any of the FI structures, and the reported energy difference is simply between the FM and the non magnetic (NM) structure. This may suggest some degree of itinerancy in the CeCo$_3$-based magnetic materials.

\begin{table}[h!]
\caption{Schematic defining the ferrimagnetic (FI) and ferromagnetic (FM) configurations of Co atoms in the Ce$_2$Co$_9$Mg primitive cell used in Curie point calculations. Sphere colors corresponds to same Co sites as in Figure~\ref{fig1}. The calculated energy difference between ($\Delta$ E) between the FI state and FM state on per Co atom basis. The calculated mean field Curie temperature ($\Delta $E/3) are also shown.}
  \centering
  \begin{tabular}{m{3cm} m{6cm}}
     \begin{minipage}{0.25\columnwidth}
    \includegraphics[width=1.3\columnwidth]{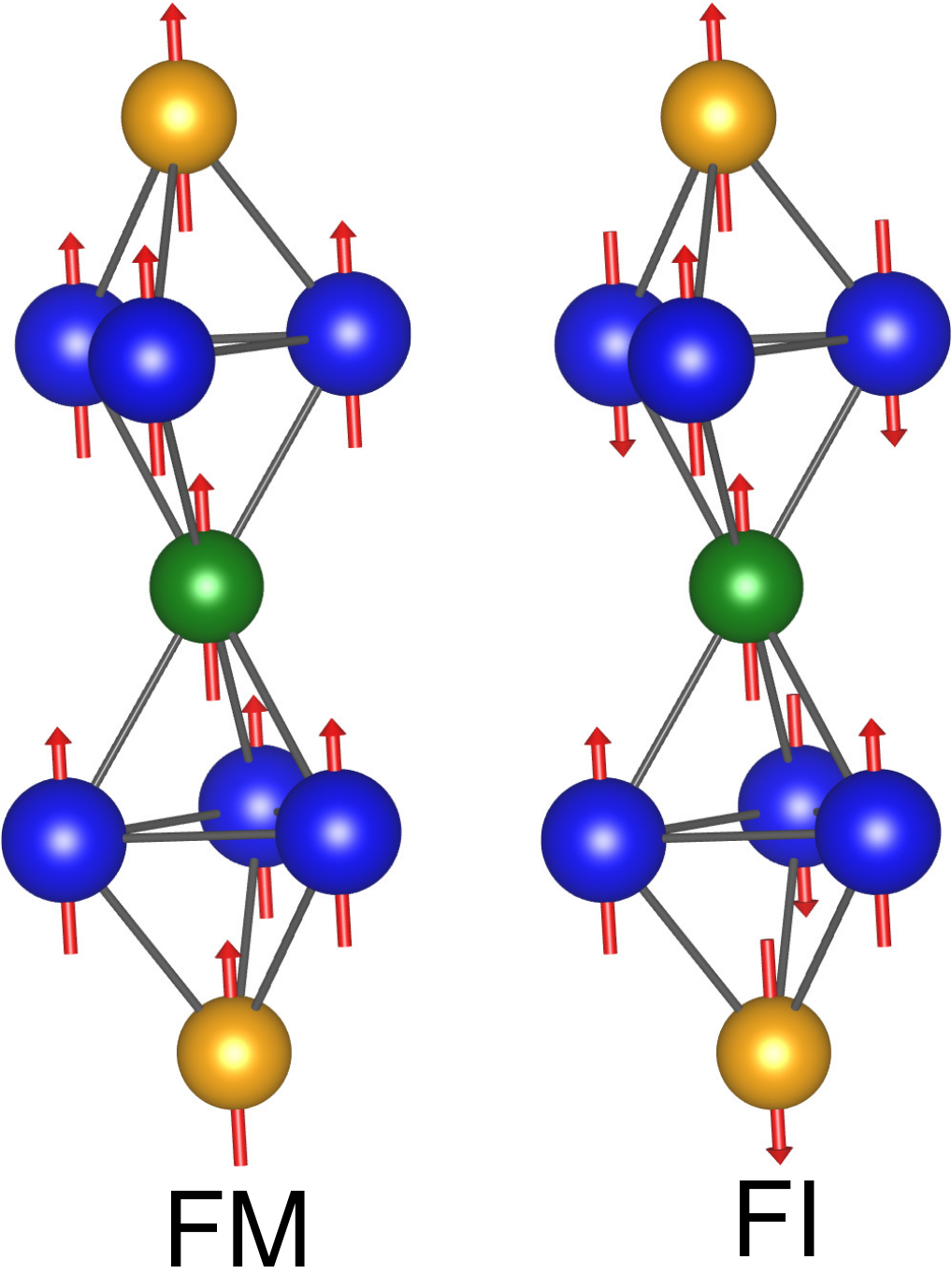}
    \end{minipage}
     &
     \begin{tabular} {c|c|c}
     \hline
		{System}  &  {$\Delta$ E (meV/Co)} & {T$_{\mathrm{C}}$} (K)\\[0.9ex]
     \hline
CeCo$_3$ & 1.9 & 7 \\[1.5ex]
Ce$_2$MgCo$_9$ & 54.8 & 212 \\[1.5ex]
CeMg$_2$Co$_9$ & 87.8 & 339 \\[1.5ex]
     \hline
     \end{tabular}
   \end{tabular}
   \label{table2}
\end{table}

As shown in Table~\ref{table2} for CeCo$_3$, the NM state is 1.9 meV/Co above the FM ground state resulting in a Curie temperature estimate of approximately 7 K. With introducing one Mg at Ce site (Ce$_2$MgCo$_9$) the energy difference between FM and FI state is 54.8 meV/Co which is nearly 27 times higher than the base compound CeCo$_3$. Finally on adding an additional Mg (CeMg$_2$Co$_9$) the difference between FM and FI states is 87.8 meV/Co, which is 44 and 1.6 times higher than $\Delta E$ for CeCo$_3$ and Ce$_2$MgCo$_9$, respectively. This analysis shows that upon Mg alloying in CeCo$_3$, the magnitude of the $\Delta E$ increases drastically. This suggests that the mean field J parameters should increase in Ce$_{3-x}$Mg$_x$Co$_9$ as \textit{x} is increased which would result in enhanced Curie temperature. This is in agreement with recent experimental results~\cite{tej-CeCo3-Mg}, which also find a manifold increase in Curie temperature with Mg alloying. It is important to mention that, generally mean field theory is not immediately applicable to magnets with some degree of itinerant character. As these CeCo$_{3}$-based magnets exhibit some degree of itinerancy, the actual theoretical values of the Curie point from a local moment approximation may be underestimated. Note that these results only provide an estimate of the ordering point, and for more accuracy Monte Carlo calculations such as atomic spin dynamics~\cite{skubic2008method,evans2014atomistic,eriksson2017atomistic} may be necessary.

\section{Conclusion}
Motivated by a recent experimental study~\cite{tej-CeCo3-Mg} we have carried out first principles calculations to understand the remarkable transformation of paramagnetic CeCo$_3$ by Mg alloying into a material with magnetic properties (including large magnetization, Curie point, and magnetic anisotropy) comparable to those of a potential permanent magnet. We find this transformation to result from a combination of Stoner physics and the magnetic anisotropy of the Ce and Co atoms.

While the specific {\it physics} of this transformation is relatively particular to this compound, the general {\it behavior} exhibited here - the metamorphosis of an initially unpromising material into a potential permanent magnet via a simple alloying strategy - is in fact an underappreciated and powerful method for the production of permanent magnet materials.  For example, the sister compound NdCo$_{3}$ \cite{tej-CeCo3-Mg,shtender2017crystal} shows a large enhancement in its Curie point from 381 K for the base compound to 633 K for Nd$_{2}$MgCo$_{9}$.
Similarly, the low Curie point (216 K \cite{jernberg1984}) of Fe$_{2}$P, which has a favorably large magnetic anisotropy of 2.3 MJ/m$^3$, increases to nearly 700 K via the simple substitution \cite{delczeg2010} of 40 atomic percent of Phosphorus by Silicon. And our own first principles calculations \cite{pandey2018b} find that alloying of the low Curie point, low magnetization ferrimagnet Fe$_2$Ta by Hafnium yields a potential permanent magnet material with a magnetic anisotropy exceeding 2 MJ/m$^{3}$.

The point of all these results is that there are likely {\it many} potential permanent magnet materials hidden in the guise of compounds that do not appear favorable for these applications, but can in fact be made so by a simple alloying strategy. In fact this approach, based on different physics, has also been used to make thermoelectric materials, based on Si-Ge alloys, sufficiently useful to power the Voyager spacecraft, despite the unfavorably large thermal conductivity of both Silicon and Germanium.  While such a strategy, applied to magnetic materials, may not ultimately yield a magnet as powerful as those based on Nd$_2$Fe$_{14}$B, it should produce a number of ``gap magnets" \cite{coey2012} , which may well fill in the substantial performance gap (measured as energy product BH$_{max}$) between non-rare-earth magnets such as Alnico and ferrite, and the rare-earth magnets SmCo$_5$ and Nd$_2$Fe$_{14}$B.  Such magnets will likely be of great utility to modern society considering the continuing worldwide industrialization and movement towards clean energy technologies such as electric vehicles and wind turbines, which often use such magnets. 

\section*{Acknowledgements}
\noindent
This research was supported by the Critical Materials Institute, an Energy Innovation Hub funded by the U.S. Department of Energy, Office of Energy Efficiency and Renewable Energy, Advanced Manufacturing Office. This research used resources of the Compute and Data Environment for Science (CADES) at the Oak Ridge National Laboratory, which is supported by the Office of Science of the U.S. Department of Energy under Contract No. DE-AC05-00OR22725. The Department of Energy will provide public access to these results of federally sponsored research in accordance with the DOE Public Access Plan (http://energy.gov/downloads/doe-public-access-plan).

\section*{Appendix}
\subsection{Effect of lattice parameters and  functional on magnetic properties of CeCo$_3$}
\begin{figure}[!ht]
\includegraphics[width=\columnwidth]{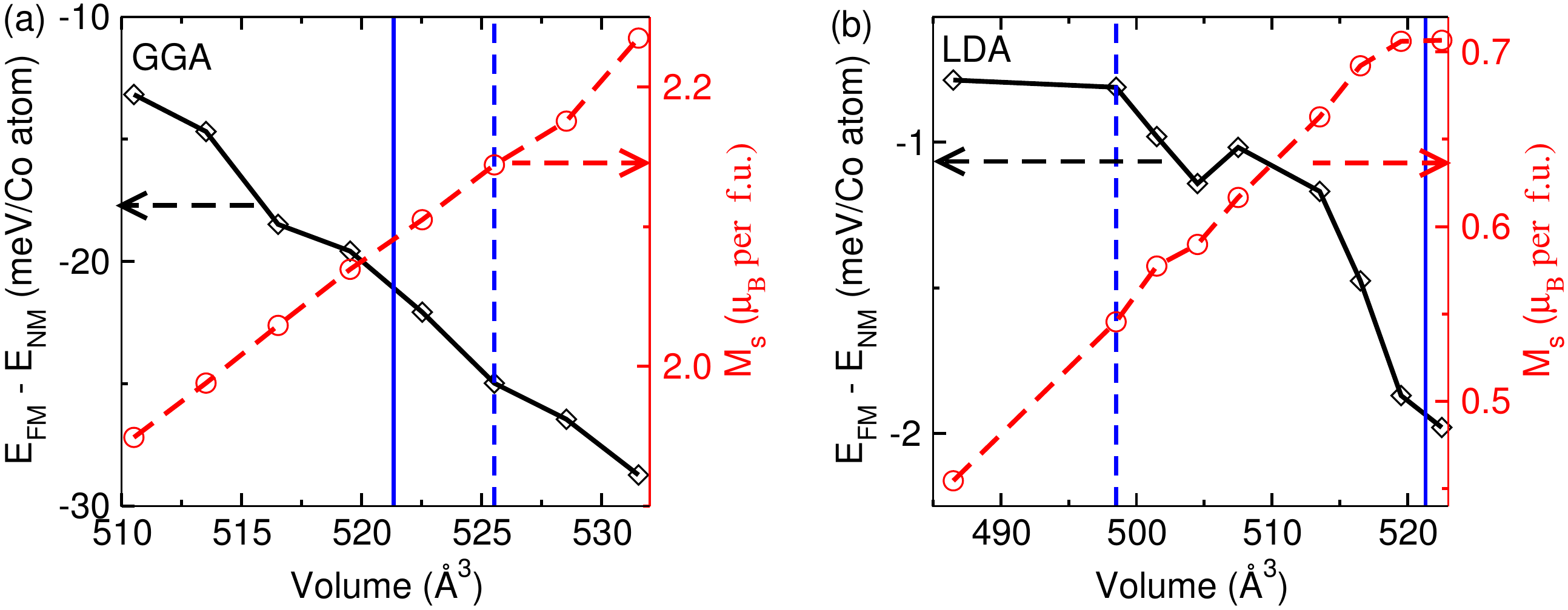}
\caption{The calculated energy difference between FM and NM state on per Co basis (left y-axis) and total unit cell magnetic moment on per formula unit basis as a function of unit cell volume for CeCo$_3$ under (a) GGA and (b) LDA  functionals. The optimized DFT volume (dashed lines) and experimental volume (solid lines), are also shown.}
\label{fig7}
\end{figure}

The reported lattice parameters for CeCo$_3$ reported within the Inorganic Crystal Structure Database (ICSD) vary by about 4 \%. Hence volume may play some role in determining the ground state of CeCo$_3$. To check the volume dependence of magnetic properties for CeCo$_3$ total energy and magnetic moments were computed as function of lattice constant within both LDA and GGA functionals. These results are shown in Figure~\ref{fig7}. Figure~\ref{fig7} shows that both LDA and GGA calculation find CeCo$_3$ to be close to a magnetic instability. While both LDA and GGA calculations favor a ferromagnetic state, at experimental lattice parameters within LDA the energy difference between ferromagnetic and non-magnetic state ($E_{\mathrm{FM}}-E_{\mathrm{NM}}$) is only $-$1.9 meV on per Co basis. On the other hand within GGA this difference is about $-$21 meV per Co atom.

For CeCo$_3$ the lattice parameters optimized within GGA (\textit{a} = 4.91~\AA, \textit{c} =  25.09~\AA) are in good agreement with experimental lattice parameters, whereas LDA underestimates the lattice parameters (\textit{a} = 4.85~\AA, \textit{c} = 24.42~\AA). Figure~\ref{fig7} (a) and (b) also shows the  behavior of the magnetic moments (plotted on right $y$ axis) in CeCo$_3$ as a function of volume. Both LDA and GGA curves show the expected increase in magnetic moments with increasing cell volume. One striking difference between LDA and GGA functional is the huge difference in calculated magnetization. At GGA optimized volume the total spin moment in the unit cell is 2.1 $\mu_B$ on per formula unit basis, whereas with LDA a relatively smaller moment of 0.55 $\mu_B$ on per formula unit basis is observed. This is agreement with previous studies on weak itinerant magnets~\cite{khmelevskyi2005magnetism, aguayo2004n,larson2004magnetism,sieberer2006magnetic} where GGA is shown to overestimate the magnetic moments. Experimentally CeCo$_3$ is often characterized as Pauli paramagnetic~\cite{buschow1980magnetic,tej-CeCo3-Mg} which is better captured in this case by the LDA functional. For Ce$_2$Co$_9$Mg both LDA and GGA functional give consistent results. At experimental lattice parameters with LDA functional we find total magnetic moment and MAE of 10.04 $\mu_B$, and 0.48 MJ/m$^3$, respectively. These are in good agreement with the total magnetic moment and MAE of 10.9 $\mu_B$ and MAE of 0.47 MJ/m$^3$ calculated within GGA. Again all the calculations presented in the main text are done within LDA.
\subsection{Determination of Stoner parameter (\textit{I})}
The exchange splitting for CeCo$_3$ depends sensitively on the \textbf{k}-vector, therefore here we calculate average exchange splitting ($\Delta E_{ex}$) which can be defined as difference of spin-up and spin-down eigenvalues in the corresponding bands~\cite{kubler2017theory,zhuang2016strong}. This gives us a value of about 0.185 eV. Now Stoner parameter can be calculated by using the relation $\Delta E_{ex} = I\mathrm{m}_{avg}$~\cite{ortenzi2011competition}. Here m$_{avg}$ is the avergae Co moment on per Co basis which is about 0.33 $\mu_B$ for CeCo$_3$. This gives us \textit{I} as 0.56 eV, which is very close to the value (0.49 eV) for elemental Co~\cite{janak1977uniform}.

\subsection{Convergence of MAE for Ce$_2$Co$_9$Mg}
Convergence of MAE with respect to number of \textbf{k}-points for Ce$_2$Co$_9$Mg is shown in Table~\ref{table3}. MAE is calculated by subtracting total energies along in-plane [1$\bar{1}$0] and out-of-plane [111] direction.  For comparison both LDA$+$SOC, and LDA$+$SOC$+$U calculations are shown. Positive value of MAE indicates uni-axial anisotropy. MAE changes by less than 4\% on going from 4000 \textbf{k}-points to 6000 \textbf{k}-point. All the calculation presented in the main text are done by using 4000 reducible \textbf{k}-points. 
\begin{table}[!ht]
\begin{center}
\caption{Convergence of MAE (in meV) with respect to the number of \textbf{k}-points in the full Brillouin zone.}
\label{table3}
\begin{tabular}{c| c| c}
\hline
No. of \textbf{k}-points  & MAE (LDA+SOC) & MAE (LDA+SOC+U) \\
\hline 
300 & 2.91 & 2.86 \\
1000 & 0.380 & 1.65\\
2000 & 0.484 & 2.16  \\
4000 &0.488 & 2.21\\
6000 & 0.473 & 2.13 \\
\hline
\end{tabular}
\end{center}
\end{table}
\subsection{Valency of Ce}
\begin{figure}[ht!]
\centering
\includegraphics[width=0.9\columnwidth]{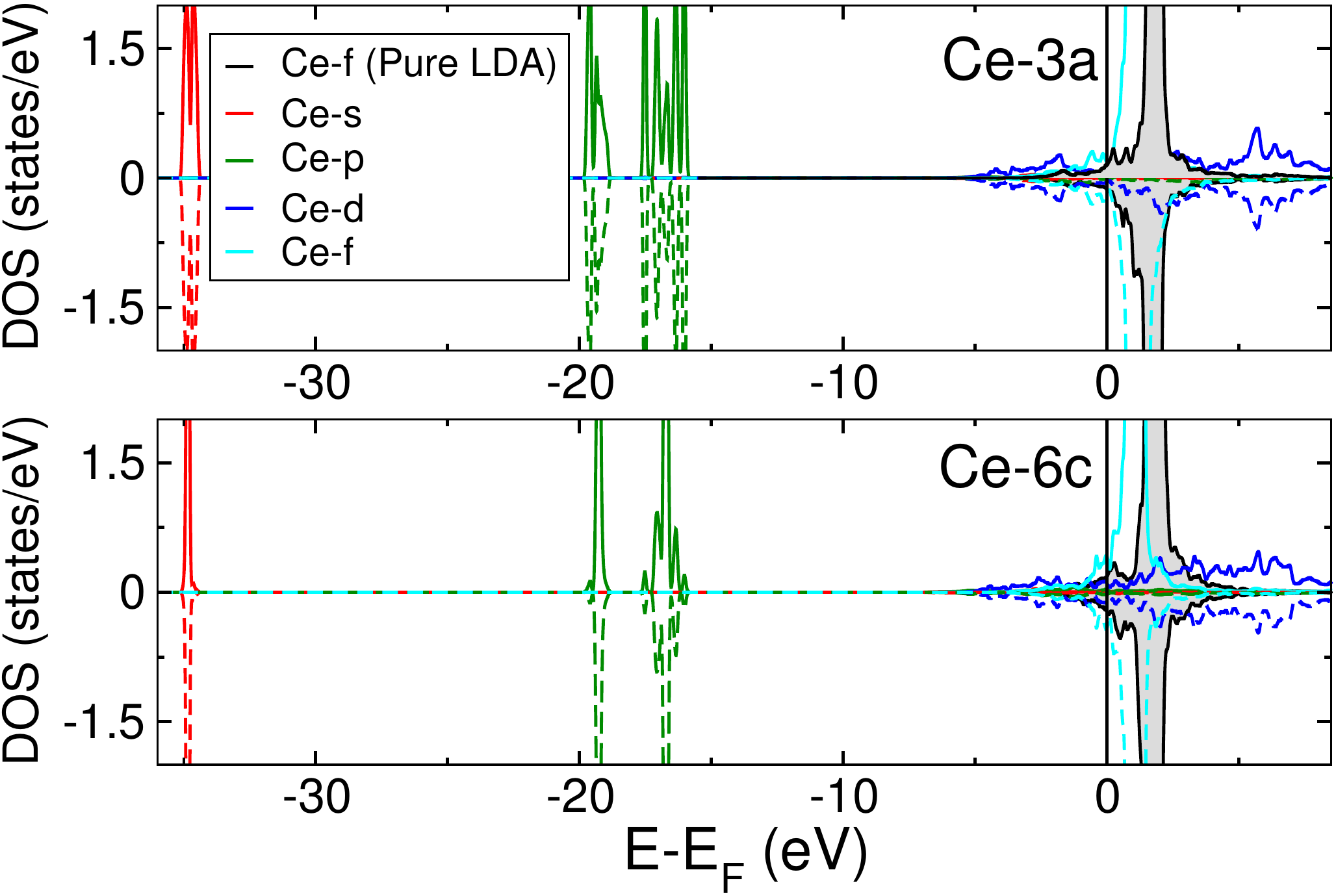}
\caption{DOS for non-equivalent Ce sites in CeCo$_3$ calculated under LDA$+$SOC$+$U with U = 1.5 eV at Ce site. DOS calculated within pure LDA (without SOC and without U) is shown by black (gray filled) lines.}
\label{fig8}
\end{figure}

The nature Ce-4$f$ valence state in Ce-transition metal complexes is still dubious. Several previous studies report that occurrence of mixed valency (Ce$^{3+}$ and Ce$^{4+}$) for Ce ion~\cite{coey1993ce2fe17,sales1975susceptibility,alam2014mixed,jarlborg2014role,korner2016theoretical}. The Ce valence can be further altered by doping~\cite{jin2016manipulating}, alloying~\cite{alam2013site,susner20172flux}, hydrogenation~\cite{isnard1998hydrogen,chaboy1995effect} etc. Our calculations find the valency of Ce in CeCo$_3$ to be dependent on the approximation used. For calculations without a U, as shown in inset of Figure~\ref{fig8} the localized Ce 4$f$ states fall above the Fermi level (filled gray line), indicating tetravalency, while for the U value of 1.5 eV used here for Ce in CeCo$_3$ some of these states fall below E$_F$ indicating trivalency. Although this does not confirm the occurrence of valence fluctuations in these compounds, the calculated magnetic properties seems to suggest possibility of mixed valency. For instant the MAE of Ce$_2$Co$_9$Mg by considering Ce-$f$ valence electrons as core states is very small (0.22 MJ/m$^3$), suggesting that Ce $f$ valence electrons are important for magnetic properties. Therefore, within our study the Ce $f$-states are included as valence states, and calculated magnetic properties are in good agreement with experiments.


\begin{thebibliography}{77}%
\makeatletter
\providecommand \@ifxundefined [1]{%
 \@ifx{#1\undefined}
}%
\providecommand \@ifnum [1]{%
 \ifnum #1\expandafter \@firstoftwo
 \else \expandafter \@secondoftwo
 \fi
}%
\providecommand \@ifx [1]{%
 \ifx #1\expandafter \@firstoftwo
 \else \expandafter \@secondoftwo
 \fi
}%
\providecommand \natexlab [1]{#1}%
\providecommand \enquote  [1]{``#1''}%
\providecommand \bibnamefont  [1]{#1}%
\providecommand \bibfnamefont [1]{#1}%
\providecommand \citenamefont [1]{#1}%
\providecommand \href@noop [0]{\@secondoftwo}%
\providecommand \href [0]{\begingroup \@sanitize@url \@href}%
\providecommand \@href[1]{\@@startlink{#1}\@@href}%
\providecommand \@@href[1]{\endgroup#1\@@endlink}%
\providecommand \@sanitize@url [0]{\catcode `\\12\catcode `\$12\catcode
  `\&12\catcode `\#12\catcode `\^12\catcode `\_12\catcode `\%12\relax}%
\providecommand \@@startlink[1]{}%
\providecommand \@@endlink[0]{}%
\providecommand \url  [0]{\begingroup\@sanitize@url \@url }%
\providecommand \@url [1]{\endgroup\@href {#1}{\urlprefix }}%
\providecommand \urlprefix  [0]{URL }%
\providecommand \Eprint [0]{\href }%
\providecommand \doibase [0]{http://dx.doi.org/}%
\providecommand \selectlanguage [0]{\@gobble}%
\providecommand \bibinfo  [0]{\@secondoftwo}%
\providecommand \bibfield  [0]{\@secondoftwo}%
\providecommand \translation [1]{[#1]}%
\providecommand \BibitemOpen [0]{}%
\providecommand \bibitemStop [0]{}%
\providecommand \bibitemNoStop [0]{.\EOS\space}%
\providecommand \EOS [0]{\spacefactor3000\relax}%
\providecommand \BibitemShut  [1]{\csname bibitem#1\endcsname}%
\let\auto@bib@innerbib\@empty
\bibitem [{\citenamefont {Mattis}\ and\ \citenamefont
  {Butrymowicz}(1965)}]{mattis1965theory}%
  \BibitemOpen
  \bibfield  {author} {\bibinfo {author} {\bibfnamefont {D.~C.}\ \bibnamefont
  {Mattis}}\ and\ \bibinfo {author} {\bibfnamefont {D.~B.}\ \bibnamefont
  {Butrymowicz}},\ }\href@noop {} The theory of magnetism, {\bibfield  {journal} {\bibinfo  {journal}
  {Physics Today}\ }\textbf {\bibinfo {volume} {18}},\ \bibinfo {pages} {66}
  (\bibinfo {year} {1965})}\BibitemShut {NoStop}%
\bibitem [{\citenamefont {Coey}(2010)}]{coey2010magnetism}%
  \BibitemOpen
  \bibfield  {author} {\bibinfo {author} {\bibfnamefont {J.~M.}\ \bibnamefont
  {Coey}},\ }\href@noop {} {\emph {\bibinfo {title} {Magnetism and magnetic
  materials}}}\ (\bibinfo  {publisher} {Cambridge University Press},\ \bibinfo
  {year} {2010})\BibitemShut {NoStop}%
\bibitem [{\citenamefont {Shull}\ \emph {et~al.}(1951)\citenamefont {Shull},
  \citenamefont {Strauser},\ and\ \citenamefont {Wollan}}]{shull1951neutron}%
  \BibitemOpen
  \bibfield  {author} {\bibinfo {author} {\bibfnamefont {C.~G.}\ \bibnamefont
  {Shull}}, \bibinfo {author} {\bibfnamefont {W.}~\bibnamefont {Strauser}}, \
  and\ \bibinfo {author} {\bibfnamefont {E.}~\bibnamefont {Wollan}},\ Neutron diffraction by
  paramagnetic and antiferromagnetic substances, 
  }\href@noop {} {\bibfield  {journal} {\bibinfo  {journal} {Phys. Rev.}\
  }\textbf {\bibinfo {volume} {83}},\ \bibinfo {pages} {333} (\bibinfo {year}
  {1951})}\BibitemShut {NoStop}%
\bibitem [{\citenamefont {Lovesey}(1984)}]{lovesey1984theory}%
  \BibitemOpen
  \bibfield  {author} {\bibinfo {author} {\bibfnamefont {S.~W.}\ \bibnamefont
  {Lovesey}},\ }\href@noop {} {\emph {\bibinfo {title} {Theory of neutron
  scattering from condensed matter}}}\ (\bibinfo  {publisher} {Oxford},\
  \bibinfo {year} {1984})\BibitemShut {NoStop}%
\bibitem [{\citenamefont {Shull}(1995)}]{shull1995early}%
  \BibitemOpen
  \bibfield  {author} {\bibinfo {author} {\bibfnamefont {C.~G.}\ \bibnamefont
  {Shull}},\ }\href@noop {} Early development of neutron scattering, {\bibfield  {journal} {\bibinfo  {journal} {Rev.
  Mod. Phys.}\ }\textbf {\bibinfo {volume} {67}},\ \bibinfo {pages} {753}
  (\bibinfo {year} {1995})}\BibitemShut {NoStop}%
\bibitem [{\citenamefont {May}\ \emph {et~al.}(2017)\citenamefont {May},
  \citenamefont {Liu}, \citenamefont {Calder}, \citenamefont {Parker},
  \citenamefont {Pandey}, \citenamefont {Cakmak}, \citenamefont {Cao},
  \citenamefont {Yan},\ and\ \citenamefont {McGuire}}]{may2017magnetic}%
  \BibitemOpen
  \bibfield  {author} {\bibinfo {author} {\bibfnamefont {A.~F.}\ \bibnamefont
  {May}}, \bibinfo {author} {\bibfnamefont {Y.}~\bibnamefont {Liu}}, \bibinfo
  {author} {\bibfnamefont {S.}~\bibnamefont {Calder}}, \bibinfo {author}
  {\bibfnamefont {D.~S.}\ \bibnamefont {Parker}}, \bibinfo {author}
  {\bibfnamefont {T.}~\bibnamefont {Pandey}}, \bibinfo {author} {\bibfnamefont
  {E.}~\bibnamefont {Cakmak}}, \bibinfo {author} {\bibfnamefont
  {H.}~\bibnamefont {Cao}}, \bibinfo {author} {\bibfnamefont {J.}~\bibnamefont
  {Yan}}, \ and\ \bibinfo {author} {\bibfnamefont {M.~A.}\ \bibnamefont
  {McGuire}},\ }\href@noop {}Magnetic order and interactions in ferrimagnetic {Mn$_3$Si$_2$Te$_6$}, {\bibfield  {journal} {\bibinfo  {journal} {Phys.
  Rev. B}\ }\textbf {\bibinfo {volume} {95}},\ \bibinfo {pages} {174440}
  (\bibinfo {year} {2017})}\BibitemShut {NoStop}%
\bibitem [{\citenamefont {May}\ \emph {et~al.}(2016)\citenamefont {May},
  \citenamefont {Calder}, \citenamefont {Parker}, \citenamefont {Sales},\ and\
  \citenamefont {McGuire}}]{may2016competing}%
  \BibitemOpen
  \bibfield  {author} {\bibinfo {author} {\bibfnamefont {A.~F.}\ \bibnamefont
  {May}}, \bibinfo {author} {\bibfnamefont {S.}~\bibnamefont {Calder}},
  \bibinfo {author} {\bibfnamefont {D.~S.}\ \bibnamefont {Parker}}, \bibinfo
  {author} {\bibfnamefont {B.~C.}\ \bibnamefont {Sales}}, \ and\ \bibinfo
  {author} {\bibfnamefont {M.~A.}\ \bibnamefont {McGuire}},\ }\href@noop {} Competing
  magnetic ground states and their coupling to the crystal lattice in {CuFe$_2$Ge$_2$},
  {\bibfield  {journal} {\bibinfo  {journal} {Sci. Rep.}\ }\textbf {\bibinfo
  {volume} {6}}, 35325 (\bibinfo {year} {2016})}\BibitemShut {NoStop}%
\bibitem [{\citenamefont {Hohenberg}\ and\ \citenamefont
  {Kohn}(1964)}]{hohenberg1964inhomogeneous}%
  \BibitemOpen
  \bibfield  {author} {\bibinfo {author} {\bibfnamefont {P.}~\bibnamefont
  {Hohenberg}}\ and\ \bibinfo {author} {\bibfnamefont {W.}~\bibnamefont
  {Kohn}},\ }\href@noop {} Inhomogeneous
  electron gas, {\bibfield  {journal} {\bibinfo  {journal} {Phys.
  Rev.}\ }\textbf {\bibinfo {volume} {136}},\ \bibinfo {pages} {B864} (\bibinfo
  {year} {1964})}\BibitemShut {NoStop}%
\bibitem [{\citenamefont {Kohn}\ and\ \citenamefont
  {Sham}(1965)}]{kohn1965self}%
  \BibitemOpen
  \bibfield  {author} {\bibinfo {author} {\bibfnamefont {W.}~\bibnamefont
  {Kohn}}\ and\ \bibinfo {author} {\bibfnamefont {L.~J.}\ \bibnamefont
  {Sham}},\ }\href@noop {} Self-consistent
  equations including exchange and correlation effects, {\bibfield  {journal} {\bibinfo  {journal} {Phys.
  Rev.}\ }\textbf {\bibinfo {volume} {140}},\ \bibinfo {pages} {A1133}
  (\bibinfo {year} {1965})}\BibitemShut {NoStop}%
\bibitem [{\citenamefont {Kohn}(1999)}]{lecture1999electronic}%
  \BibitemOpen
  \bibfield  {author} {\bibinfo {author} {\bibfnamefont {W.}~\bibnamefont
  {Kohn}},\ }\href@noop {} Electronic
  structure of matter-wave functions and density functionals, {\bibfield  {journal} {\bibinfo  {journal} {Rev.
  Mod. Phys.}\ }\textbf {\bibinfo {volume} {71}},\ \bibinfo {pages} {1253}
  (\bibinfo {year} {1999})}\BibitemShut {NoStop}%
\bibitem [{\citenamefont {Kohn}\ \emph {et~al.}(1996)\citenamefont {Kohn},
  \citenamefont {Becke},\ and\ \citenamefont {Parr}}]{kohn1996density}%
  \BibitemOpen
  \bibfield  {author} {\bibinfo {author} {\bibfnamefont {W.}~\bibnamefont
  {Kohn}}, \bibinfo {author} {\bibfnamefont {A.~D.}\ \bibnamefont {Becke}}, \
  and\ \bibinfo {author} {\bibfnamefont {R.~G.}\ \bibnamefont {Parr}},\
  }\href@noop {} Density functional theory of
  electronic structure, {\bibfield  {journal} {\bibinfo  {journal} {J. Phys. Chem.}\
  }\textbf {\bibinfo {volume} {100}},\ \bibinfo {pages} {12974} (\bibinfo
  {year} {1996})}\BibitemShut {NoStop}%
\bibitem [{\citenamefont {Perdew}\ \emph {et~al.}(1992)\citenamefont {Perdew},
  \citenamefont {Chevary}, \citenamefont {Vosko}, \citenamefont {Jackson},
  \citenamefont {Pederson}, \citenamefont {Singh},\ and\ \citenamefont
  {Fiolhais}}]{perdew1992atoms}%
  \BibitemOpen
  \bibfield  {author} {\bibinfo {author} {\bibfnamefont {J.~P.}\ \bibnamefont
  {Perdew}}, \bibinfo {author} {\bibfnamefont {J.~A.}\ \bibnamefont {Chevary}},
  \bibinfo {author} {\bibfnamefont {S.~H.}\ \bibnamefont {Vosko}}, \bibinfo
  {author} {\bibfnamefont {K.~A.}\ \bibnamefont {Jackson}}, \bibinfo {author}
  {\bibfnamefont {M.~R.}\ \bibnamefont {Pederson}}, \bibinfo {author}
  {\bibfnamefont {D.~J.}\ \bibnamefont {Singh}}, \ and\ \bibinfo {author}
  {\bibfnamefont {C.}~\bibnamefont {Fiolhais}},\ }\href@noop {}Atoms, molecules, solids, and surfaces:
  Applications of the generalized gradient approximation for exchange and
  correlation, {\bibfield
  {journal} {\bibinfo  {journal} {Phys. Rev. B}\ }\textbf {\bibinfo {volume}
  {46}},\ \bibinfo {pages} {6671} (\bibinfo {year} {1992})}\BibitemShut
  {NoStop}%
\bibitem [{\citenamefont {Perdew}\ \emph {et~al.}(2005)\citenamefont {Perdew},
  \citenamefont {Ruzsinszky}, \citenamefont {Tao}, \citenamefont {Staroverov},
  \citenamefont {Scuseria},\ and\ \citenamefont
  {Csonka}}]{perdew2005prescription}%
  \BibitemOpen
  \bibfield  {author} {\bibinfo {author} {\bibfnamefont {J.~P.}\ \bibnamefont
  {Perdew}}, \bibinfo {author} {\bibfnamefont {A.}~\bibnamefont {Ruzsinszky}},
  \bibinfo {author} {\bibfnamefont {J.}~\bibnamefont {Tao}}, \bibinfo {author}
  {\bibfnamefont {V.~N.}\ \bibnamefont {Staroverov}}, \bibinfo {author}
  {\bibfnamefont {G.~E.}\ \bibnamefont {Scuseria}}, \ and\ \bibinfo {author}
  {\bibfnamefont {G.~I.}\ \bibnamefont {Csonka}},\ }\href@noop {}Prescription for the design and selection of density functional
  approximations: More constraint satisfaction with fewer fits, {\bibfield
  {journal} {\bibinfo  {journal} {J. Chem. Phys.}\ }\textbf {\bibinfo {volume}
  {123}},\ \bibinfo {pages} {062201} (\bibinfo {year} {2005})}\BibitemShut
  {NoStop}%
\bibitem [{\citenamefont {Perdew}\ \emph {et~al.}(1996)\citenamefont {Perdew},
  \citenamefont {Burke},\ and\ \citenamefont
  {Ernzerhof}}]{perdew1996generalized}%
  \BibitemOpen
  \bibfield  {author} {\bibinfo {author} {\bibfnamefont {J.~P.}\ \bibnamefont
  {Perdew}}, \bibinfo {author} {\bibfnamefont {K.}~\bibnamefont {Burke}}, \
  and\ \bibinfo {author} {\bibfnamefont {M.}~\bibnamefont {Ernzerhof}},\
  }\href@noop {}Generalized
  gradient approximation made simple, {\bibfield  {journal} {\bibinfo  {journal} {Phys. Rev. Lett.}\ }\textbf {\bibinfo {volume} {77}},\ \bibinfo {pages} {3865}
  (\bibinfo {year} {1996})}\BibitemShut {NoStop}%
\bibitem [{\citenamefont {Kresse}\ and\ \citenamefont
  {Furthm{\"u}ller}(1996)}]{kresse1996efficient}%
  \BibitemOpen
  \bibfield  {author} {\bibinfo {author} {\bibfnamefont {G.}~\bibnamefont
  {Kresse}}\ and\ \bibinfo {author} {\bibfnamefont {J.}~\bibnamefont
  {Furthm{\"u}ller}},\ }\href@noop {}Efficient iterative schemes for ab initio total-energy calculations using a
  plane-wave basis set, {\bibfield  {journal} {\bibinfo
  {journal} {Phys. Rev. B}\ }\textbf {\bibinfo {volume} {54}},\ \bibinfo
  {pages} {11169} (\bibinfo {year} {1996})}\BibitemShut {NoStop}%
\bibitem [{\citenamefont {Singh}\ and\ \citenamefont
  {Nordstrom}(2006)}]{singh2006planewaves}%
  \BibitemOpen
  \bibfield  {author} {\bibinfo {author} {\bibfnamefont {D.~J.}\ \bibnamefont
  {Singh}}\ and\ \bibinfo {author} {\bibfnamefont {L.}~\bibnamefont
  {Nordstrom}},\ }\href@noop {} {\emph {\bibinfo {title} {Planewaves,
  Pseudopotentials, and the LAPW method}}}\ (\bibinfo  {publisher} {Springer
  Science \& Business Media},\ \bibinfo {year} {2006})\BibitemShut {NoStop}%
\bibitem [{\citenamefont {Drebov}\ \emph {et~al.}(2013)\citenamefont {Drebov},
  \citenamefont {Martinez-Limia}, \citenamefont {Kunz}, \citenamefont {Gola},
  \citenamefont {Shigematsu}, \citenamefont {Eckl}, \citenamefont {Gumbsch},\
  and\ \citenamefont {Els{\"a}sser}}]{drebov2013ab}%
  \BibitemOpen
  \bibfield  {author} {\bibinfo {author} {\bibfnamefont {N.}~\bibnamefont
  {Drebov}}, \bibinfo {author} {\bibfnamefont {A.}~\bibnamefont
  {Martinez-Limia}}, \bibinfo {author} {\bibfnamefont {L.}~\bibnamefont
  {Kunz}}, \bibinfo {author} {\bibfnamefont {A.}~\bibnamefont {Gola}}, \bibinfo
  {author} {\bibfnamefont {T.}~\bibnamefont {Shigematsu}}, \bibinfo {author}
  {\bibfnamefont {T.}~\bibnamefont {Eckl}}, \bibinfo {author} {\bibfnamefont
  {P.}~\bibnamefont {Gumbsch}}, \ and\ \bibinfo {author} {\bibfnamefont
  {C.}~\bibnamefont {Els{\"a}sser}},\ }\href@noop {}Ab initio screening
  methodology applied to the search for new permanent magnetic materials, {\bibfield  {journal}
  {\bibinfo  {journal} {New J. Phys.}\ }\textbf {\bibinfo {volume} {15}},\
  \bibinfo {pages} {125023} (\bibinfo {year} {2013})}\BibitemShut {NoStop}%
\bibitem [{\citenamefont {K{\"o}rner}\ \emph {et~al.}(2016)\citenamefont
  {K{\"o}rner}, \citenamefont {Krugel},\ and\ \citenamefont
  {Els{\"a}sser}}]{korner2016theoretical}%
  \BibitemOpen
  \bibfield  {author} {\bibinfo {author} {\bibfnamefont {W.}~\bibnamefont
  {K{\"o}rner}}, \bibinfo {author} {\bibfnamefont {G.}~\bibnamefont {Krugel}},
  \ and\ \bibinfo {author} {\bibfnamefont {C.}~\bibnamefont {Els{\"a}sser}},\
  }\href@noop {}Theoretical screening of intermetallic {ThMn$_{12}$}-type phases for new
  hard-magnetic compounds with low rare earth content, {\bibfield  {journal} {\bibinfo  {journal} {Sci. Rep.}\
  }\textbf {\bibinfo {volume} {6}},\ \bibinfo {pages} {24686} (\bibinfo
  {year} {2016})}\BibitemShut {NoStop}%
\bibitem [{\citenamefont {Sanvito}\ \emph {et~al.}(2017)\citenamefont
  {Sanvito}, \citenamefont {Oses}, \citenamefont {Xue}, \citenamefont {Tiwari},
  \citenamefont {Zic}, \citenamefont {Archer}, \citenamefont {Tozman},
  \citenamefont {Venkatesan}, \citenamefont {Coey},\ and\ \citenamefont
  {Curtarolo}}]{sanvito2017accelerated}%
  \BibitemOpen
  \bibfield  {author} {\bibinfo {author} {\bibfnamefont {S.}~\bibnamefont
  {Sanvito}}, \bibinfo {author} {\bibfnamefont {C.}~\bibnamefont {Oses}},
  \bibinfo {author} {\bibfnamefont {J.}~\bibnamefont {Xue}}, \bibinfo {author}
  {\bibfnamefont {A.}~\bibnamefont {Tiwari}}, \bibinfo {author} {\bibfnamefont
  {M.}~\bibnamefont {Zic}}, \bibinfo {author} {\bibfnamefont {T.}~\bibnamefont
  {Archer}}, \bibinfo {author} {\bibfnamefont {P.}~\bibnamefont {Tozman}},
  \bibinfo {author} {\bibfnamefont {M.}~\bibnamefont {Venkatesan}}, \bibinfo
  {author} {\bibfnamefont {M.}~\bibnamefont {Coey}}, \ and\ \bibinfo {author}
  {\bibfnamefont {S.}~\bibnamefont {Curtarolo}},\ }\href@noop {}Accelerated discovery of new magnets in the heusler alloy family, {\bibfield
  {journal} {\bibinfo  {journal} {Sci. Adv.}\ }\textbf {\bibinfo {volume}
  {3}},\ \bibinfo {pages} {e1602241} (\bibinfo {year} {2017})}\BibitemShut
  {NoStop}%
\bibitem [{\citenamefont {Curtarolo}\ \emph {et~al.}(2012)\citenamefont
  {Curtarolo}, \citenamefont {Setyawan}, \citenamefont {Wang}, \citenamefont
  {Xue}, \citenamefont {Yang}, \citenamefont {Taylor}, \citenamefont {Nelson},
  \citenamefont {Hart}, \citenamefont {Sanvito}, \citenamefont
  {Buongiorno-Nardelli} \emph {et~al.}}]{curtarolo2012aflowlib}%
  \BibitemOpen
  \bibfield  {author} {\bibinfo {author} {\bibfnamefont {S.}~\bibnamefont
  {Curtarolo}}, \bibinfo {author} {\bibfnamefont {W.}~\bibnamefont {Setyawan}},
  \bibinfo {author} {\bibfnamefont {S.}~\bibnamefont {Wang}}, \bibinfo {author}
  {\bibfnamefont {J.}~\bibnamefont {Xue}}, \bibinfo {author} {\bibfnamefont
  {K.}~\bibnamefont {Yang}}, \bibinfo {author} {\bibfnamefont {R.~H.}\
  \bibnamefont {Taylor}}, \bibinfo {author} {\bibfnamefont {L.~J.}\
  \bibnamefont {Nelson}}, \bibinfo {author} {\bibfnamefont {G.~L.}\
  \bibnamefont {Hart}}, \bibinfo {author} {\bibfnamefont {S.}~\bibnamefont
  {Sanvito}}, \bibinfo {author} {\bibfnamefont {M.}~\bibnamefont
  {Buongiorno-Nardelli}},  \emph {et~al.},\ }\href@noop {}{Aflowlib}. org: A
  distributed materials properties repository from high-throughput ab initio
  calculations, {\bibfield
  {journal} {\bibinfo  {journal} {Comput. Mater. Sci.}\ }\textbf {\bibinfo
  {volume} {58}},\ \bibinfo {pages} {227} (\bibinfo {year} {2012})}\BibitemShut
  {NoStop}%
\bibitem [{\citenamefont {Coey}(1986)}]{coey1986intrinsic}%
  \BibitemOpen
  \bibfield  {author} {\bibinfo {author} {\bibfnamefont {J.}~\bibnamefont
  {Coey}},\ }\href@noop {}Intrinsic magnetic
  properties of compounds with the {Nd$_2$Fe$_{14}$B} structure, {\bibfield  {journal} {\bibinfo  {journal} {J. Less
  Common Met.}\ }\textbf {\bibinfo {volume} {126}},\ \bibinfo {pages} {21}
  (\bibinfo {year} {1986})}\BibitemShut {NoStop}%
\bibitem [{\citenamefont {Susner}\ \emph {et~al.}(2017)\citenamefont {Susner},
  \citenamefont {Conner}, \citenamefont {Saparov}, \citenamefont {McGuire},
  \citenamefont {Crumlin}, \citenamefont {Veith}, \citenamefont {Cao},
  \citenamefont {Shanavas}, \citenamefont {Parker}, \citenamefont {Chakoumakos}
  \emph {et~al.}}]{susner20172flux}%
  \BibitemOpen
  \bibfield  {author} {\bibinfo {author} {\bibfnamefont {M.~A.}\ \bibnamefont
  {Susner}}, \bibinfo {author} {\bibfnamefont {B.~S.}\ \bibnamefont {Conner}},
  \bibinfo {author} {\bibfnamefont {B.~I.}\ \bibnamefont {Saparov}}, \bibinfo
  {author} {\bibfnamefont {M.~A.}\ \bibnamefont {McGuire}}, \bibinfo {author}
  {\bibfnamefont {E.~J.}\ \bibnamefont {Crumlin}}, \bibinfo {author}
  {\bibfnamefont {G.~M.}\ \bibnamefont {Veith}}, \bibinfo {author}
  {\bibfnamefont {H.}~\bibnamefont {Cao}}, \bibinfo {author} {\bibfnamefont
  {K.~V.}\ \bibnamefont {Shanavas}}, \bibinfo {author} {\bibfnamefont {D.~S.}\
  \bibnamefont {Parker}}, \bibinfo {author} {\bibfnamefont {B.~C.}\
  \bibnamefont {Chakoumakos}},  \emph {et~al.},\ }\href@noop {}{2 flux growth and characterization of Ce-substituted Nd$_{2}$Fe$_{14}$B single
  crystals} {\bibfield
  {journal} {\bibinfo  {journal} {J. Magn. Magn. Mater.}\ }\textbf {\bibinfo
  {volume} {434}},\ \bibinfo {pages} {1} (\bibinfo {year} {2017})}\BibitemShut
  {NoStop}%
\bibitem [{\citenamefont {Tatetsu}\ \emph {et~al.}(2016)\citenamefont
  {Tatetsu}, \citenamefont {Tsuneyuki},\ and\ \citenamefont
  {Gohda}}]{tatetsu2016first}%
  \BibitemOpen
  \bibfield  {author} {\bibinfo {author} {\bibfnamefont {Y.}~\bibnamefont
  {Tatetsu}}, \bibinfo {author} {\bibfnamefont {S.}~\bibnamefont {Tsuneyuki}},
  \ and\ \bibinfo {author} {\bibfnamefont {Y.}~\bibnamefont {Gohda}},\
  }\href@noop {} {First-principles study of the role of Cu
  in improving the coercivity of Nd-Fe-B permanent magnets}, {\bibfield  {journal} {\bibinfo  {journal} {Phys. Rev.
  Appl.}\ }\textbf {\bibinfo {volume} {6}},\ \bibinfo {pages} {064029}
  (\bibinfo {year} {2016})}\BibitemShut {NoStop}%
\bibitem [{\citenamefont {Yi}\ \emph {et~al.}(2017)\citenamefont {Yi},
  \citenamefont {Zhang}, \citenamefont {Gutfleisch},\ and\ \citenamefont
  {Xu}}]{yi2017multiscale}%
  \BibitemOpen
  \bibfield  {author} {\bibinfo {author} {\bibfnamefont {M.}~\bibnamefont
  {Yi}}, \bibinfo {author} {\bibfnamefont {H.}~\bibnamefont {Zhang}}, \bibinfo
  {author} {\bibfnamefont {O.}~\bibnamefont {Gutfleisch}}, \ and\ \bibinfo
  {author} {\bibfnamefont {B.-X.}\ \bibnamefont {Xu}},\ }\href@noop {}{Multiscale examination of strain
  effects in Nd-Fe-B permanent magnets},
  {\bibfield  {journal} {\bibinfo  {journal} {Phys. Rev. Appl.}\
  }\textbf {\bibinfo {volume} {8}},\ \bibinfo {pages} {014011} (\bibinfo {year}
  {2017})}\BibitemShut {NoStop}%
\bibitem [{\citenamefont {Lamichhane}\ \emph {et~al.}(2018)\citenamefont
  {Lamichhane}, \citenamefont {Taufour}, \citenamefont {Palasyuk},
  \citenamefont {Lin}, \citenamefont {Bud’ko},\ and\ \citenamefont
  {Canfield}}]{tej-CeCo3-Mg}%
  \BibitemOpen
  \bibfield  {author} {\bibinfo {author} {\bibfnamefont {T.~N.}\ \bibnamefont
  {Lamichhane}}, \bibinfo {author} {\bibfnamefont {V.}~\bibnamefont {Taufour}},
  \bibinfo {author} {\bibfnamefont {A.}~\bibnamefont {Palasyuk}}, \bibinfo
  {author} {\bibfnamefont {Q.}~\bibnamefont {Lin}}, \bibinfo {author}
  {\bibfnamefont {S.~L.}\ \bibnamefont {Bud’ko}}, \ and\ \bibinfo {author}
  {\bibfnamefont {P.~C.}\ \bibnamefont {Canfield}},\ }\href@noop {}{Ce$_{3-x}$Mg$_x$Co$_9$: Transformation of a pauli paramagnet into a strong
  permanent magnet}, {\bibfield
  {journal} {\bibinfo  {journal} {Phys. Rev. Appl.}\ }\textbf {\bibinfo
  {volume} {9}},\ \bibinfo {pages} {024023} (\bibinfo {year}
  {2018})}\BibitemShut {NoStop}%
\bibitem [{\citenamefont {Heller}(1967)}]{heller1967experimental}%
  \BibitemOpen
  \bibfield  {author} {\bibinfo {author} {\bibfnamefont {P.}~\bibnamefont
  {Heller}},\ }\href@noop {}Experimental
  investigations of critical phenomena, {\bibfield  {journal} {\bibinfo  {journal} {Rep.
  Prog. Phys.}\ }\textbf {\bibinfo {volume} {30}},\ \bibinfo {pages} {731}
  (\bibinfo {year} {1967})}\BibitemShut {NoStop}%
\bibitem [{\citenamefont {Keffer}(1966)}]{keffer1966handbuch}%
  \BibitemOpen
  \bibfield  {author} {\bibinfo {author} {\bibfnamefont {F.}~\bibnamefont
  {Keffer}},\ }\href@noop {} {\enquote {\bibinfo {title} {Handbuch der physik,
  vol. 18-2},}\ } (\bibinfo {year} {1966})\BibitemShut {NoStop}%
\bibitem [{\citenamefont {Shtender}\ \emph {et~al.}(2017)\citenamefont
  {Shtender}, \citenamefont {Denys}, \citenamefont {Paul-Boncour},
  \citenamefont {Zavaliy}, \citenamefont {Verbovytskyy},\ and\ \citenamefont
  {Taylor}}]{shtender2017crystal}%
  \BibitemOpen
  \bibfield  {author} {\bibinfo {author} {\bibfnamefont {V.}~\bibnamefont
  {Shtender}}, \bibinfo {author} {\bibfnamefont {R.}~\bibnamefont {Denys}},
  \bibinfo {author} {\bibfnamefont {V.}~\bibnamefont {Paul-Boncour}}, \bibinfo
  {author} {\bibfnamefont {I.~Y.}\ \bibnamefont {Zavaliy}}, \bibinfo {author}
  {\bibfnamefont {Y.~V.}\ \bibnamefont {Verbovytskyy}}, \ and\ \bibinfo
  {author} {\bibfnamefont {D.}~\bibnamefont {Taylor}},\ }\href@noop {}{Crystal structure, hydrogen absorption-desorption behavior
  and magnetic properties of the Nd$_{3-x}$Mg$_x$Co$_9$ alloys},
  {\bibfield  {journal} {\bibinfo  {journal} {J. Alloys Compd.}\ }\textbf
  {\bibinfo {volume} {695}},\ \bibinfo {pages} {1426} (\bibinfo {year}
  {2017})}\BibitemShut {NoStop}%
\bibitem [{wie()}]{wien2k}%
  \BibitemOpen
  \href@noop {} {\emph {\bibinfo {title} {{WIEN2K}, {A}n {A}ugmented {P}lane
  {W}ave + {L}ocal {O}rbitals {P}rogram for {C}alculating {C}rystal
  {P}roperties}}}\BibitemShut {NoStop}%
\bibitem [{\citenamefont {Brooks}\ \emph {et~al.}(1991)\citenamefont {Brooks},
  \citenamefont {Nordstrom},\ and\ \citenamefont
  {Johansson}}]{brooks1991origin}%
  \BibitemOpen
  \bibfield  {author} {\bibinfo {author} {\bibfnamefont {M.}~\bibnamefont
  {Brooks}}, \bibinfo {author} {\bibfnamefont {L.}~\bibnamefont {Nordstrom}}, \
  and\ \bibinfo {author} {\bibfnamefont {B.}~\bibnamefont {Johansson}},\
  }\href@noop {}{Origin and ab initio
  evaluation of magnetic interactions in rare earth intermetallics}, {\bibfield  {journal} {\bibinfo  {journal} {J. Phys. Condens. Matter}\ }\textbf {\bibinfo {volume} {3}},\ \bibinfo
  {pages} {3393} (\bibinfo {year} {1991})}\BibitemShut {NoStop}%
\bibitem [{\citenamefont {Larson}\ \emph {et~al.}(2003)\citenamefont {Larson},
  \citenamefont {Mazin},\ and\ \citenamefont
  {Papaconstantopoulos}}]{larson2003calculation}%
  \BibitemOpen
  \bibfield  {author} {\bibinfo {author} {\bibfnamefont {P.}~\bibnamefont
  {Larson}}, \bibinfo {author} {\bibfnamefont {I.}~\bibnamefont {Mazin}}, \
  and\ \bibinfo {author} {\bibfnamefont {D.}~\bibnamefont
  {Papaconstantopoulos}},\ }\href@noop {} {Calculation of magnetic anisotropy energy in SmCo$_5$}, {\bibfield  {journal} {\bibinfo
  {journal} {Phys. Rev. B}\ }\textbf {\bibinfo {volume} {67}},\ \bibinfo
  {pages} {214405} (\bibinfo {year} {2003})}\BibitemShut {NoStop}%
\bibitem [{\citenamefont {Larson}\ and\ \citenamefont
  {Mazin}(2003{\natexlab{a}})}]{larson2003calculation1}%
  \BibitemOpen
  \bibfield  {author} {\bibinfo {author} {\bibfnamefont {P.}~\bibnamefont
  {Larson}}\ and\ \bibinfo {author} {\bibfnamefont {I.}~\bibnamefont {Mazin}},\
  }\href@noop {}{Calculation of magnetic
  anisotropy energy in YCo$_5$}, {\bibfield  {journal} {\bibinfo  {journal} {J. Magn. Magn. Mater.}\ }\textbf {\bibinfo {volume} {264}},\
  \bibinfo {pages} {7} (\bibinfo {year} {2003}{\natexlab{a}})}\BibitemShut
  {NoStop}%
\bibitem [{\citenamefont {Anisimov}\ \emph {et~al.}(1993)\citenamefont
  {Anisimov}, \citenamefont {Solovyev}, \citenamefont {Korotin}, \citenamefont
  {Czy{\.z}yk},\ and\ \citenamefont {Sawatzky}}]{anisimov1993density}%
  \BibitemOpen
  \bibfield  {author} {\bibinfo {author} {\bibfnamefont {V.~I.}\ \bibnamefont
  {Anisimov}}, \bibinfo {author} {\bibfnamefont {I.}~\bibnamefont {Solovyev}},
  \bibinfo {author} {\bibfnamefont {M.}~\bibnamefont {Korotin}}, \bibinfo
  {author} {\bibfnamefont {M.}~\bibnamefont {Czy{\.z}yk}}, \ and\ \bibinfo
  {author} {\bibfnamefont {G.}~\bibnamefont {Sawatzky}},\ }\href@noop {}{Density-functional theory and NiO
  photoemission spectra},
  {\bibfield  {journal} {\bibinfo  {journal} {Phys. Rev. B}\ }\textbf {\bibinfo
  {volume} {48}},\ \bibinfo {pages} {16929} (\bibinfo {year}
  {1993})}\BibitemShut {NoStop}%
\bibitem [{\citenamefont {Liechtenstein}\ \emph {et~al.}(1995)\citenamefont
  {Liechtenstein}, \citenamefont {Anisimov},\ and\ \citenamefont
  {Zaanen}}]{liechtenstein1995density}%
  \BibitemOpen
  \bibfield  {author} {\bibinfo {author} {\bibfnamefont {A.}~\bibnamefont
  {Liechtenstein}}, \bibinfo {author} {\bibfnamefont {V.}~\bibnamefont
  {Anisimov}}, \ and\ \bibinfo {author} {\bibfnamefont {J.}~\bibnamefont
  {Zaanen}},\ }\href@noop {}{Density-functional theory and strong interactions: Orbital ordering in
  Mott-Hubbard insulators}, {\bibfield  {journal} {\bibinfo  {journal} {Phys.
  Rev. B}\ }\textbf {\bibinfo {volume} {52}},\ \bibinfo {pages} {R5467}
  (\bibinfo {year} {1995})}\BibitemShut {NoStop}%
\bibitem [{\citenamefont {Shick}\ \emph {et~al.}(1999)\citenamefont {Shick},
  \citenamefont {Liechtenstein},\ and\ \citenamefont
  {Pickett}}]{shick1999implementation}%
  \BibitemOpen
  \bibfield  {author} {\bibinfo {author} {\bibfnamefont {A.}~\bibnamefont
  {Shick}}, \bibinfo {author} {\bibfnamefont {A.}~\bibnamefont
  {Liechtenstein}}, \ and\ \bibinfo {author} {\bibfnamefont {W.}~\bibnamefont
  {Pickett}},\ }\href@noop {}{Implementation
  of the LDA$+$U method using the full-potential linearized augmented plane-wave
  basis}, {\bibfield  {journal} {\bibinfo  {journal} {Phys.
  Rev. B}\ }\textbf {\bibinfo {volume} {60}},\ \bibinfo {pages} {10763}
  (\bibinfo {year} {1999})}\BibitemShut {NoStop}%
\bibitem [{\citenamefont {Madsen}\ and\ \citenamefont
  {Nov{\'a}k}(2005)}]{madsen2005charge}%
  \BibitemOpen
  \bibfield  {author} {\bibinfo {author} {\bibfnamefont {G.~K.}\ \bibnamefont
  {Madsen}}\ and\ \bibinfo {author} {\bibfnamefont {P.}~\bibnamefont
  {Nov{\'a}k}},\ }\href@noop {} {Charge order in magnetite. an LDA$+$U study}, {\bibfield  {journal} {\bibinfo  {journal}
  {Europhys. Lett.}\ }\textbf {\bibinfo {volume} {69}},\ \bibinfo {pages} {777}
  (\bibinfo {year} {2005})}\BibitemShut {NoStop}%
\bibitem [{\citenamefont {Mazin}\ \emph {et~al.}(2004)\citenamefont {Mazin},
  \citenamefont {Singh},\ and\ \citenamefont {Aguayo}}]{mazin2004density}%
  \BibitemOpen
  \bibfield  {author} {\bibinfo {author} {\bibfnamefont {I.}~\bibnamefont
  {Mazin}}, \bibinfo {author} {\bibfnamefont {D.}~\bibnamefont {Singh}}, \ and\
  \bibinfo {author} {\bibfnamefont {A.}~\bibnamefont {Aguayo}},\ }in\
  \href@noop {} {\emph {\bibinfo {booktitle} {Physics of Spin in Solids:
  Materials, Methods and Applications}}}\ (\bibinfo  {publisher} {Springer},\
  \bibinfo {year} {2004})\ pp.\ \bibinfo {pages} {139--154}\BibitemShut
  {NoStop}%
\bibitem [{\citenamefont {Khmelevskyi}\ \emph {et~al.}(2005)\citenamefont
  {Khmelevskyi}, \citenamefont {Mohn}, \citenamefont {Redinger},\ and\
  \citenamefont {Weinert}}]{khmelevskyi2005magnetism}%
  \BibitemOpen
  \bibfield  {author} {\bibinfo {author} {\bibfnamefont {S.}~\bibnamefont
  {Khmelevskyi}}, \bibinfo {author} {\bibfnamefont {P.}~\bibnamefont {Mohn}},
  \bibinfo {author} {\bibfnamefont {J.}~\bibnamefont {Redinger}}, \ and\
  \bibinfo {author} {\bibfnamefont {M.}~\bibnamefont {Weinert}},\ }\href@noop {}{Magnetism on the surface of
  the bulk paramagnetic intermetallic compound YCo$_2$}, {\bibfield  {journal} {\bibinfo  {journal} {Phys. Rev. Lett.}\
  }\textbf {\bibinfo {volume} {94}},\ \bibinfo {pages} {146403} (\bibinfo
  {year} {2005})}\BibitemShut {NoStop}%
\bibitem [{\citenamefont {Aguayo}\ \emph {et~al.}(2004)\citenamefont {Aguayo},
  \citenamefont {Mazin},\ and\ \citenamefont {Singh}}]{aguayo2004n}%
  \BibitemOpen
  \bibfield  {author} {\bibinfo {author} {\bibfnamefont {A.}~\bibnamefont
  {Aguayo}}, \bibinfo {author} {\bibfnamefont {I.}~\bibnamefont {Mazin}}, \
  and\ \bibinfo {author} {\bibfnamefont {D.}~\bibnamefont {Singh}},\
  }\href@noop {}{Why Ni$_3$Al is an itinerant
  ferromagnet but Ni$_3$Ga is not}, {\bibfield  {journal} {\bibinfo  {journal} {Phys. Rev. Lett.}\ }\textbf {\bibinfo {volume} {92}},\ \bibinfo {pages} {147201}
  (\bibinfo {year} {2004})}\BibitemShut {NoStop}%
\bibitem [{\citenamefont {Larson}\ \emph {et~al.}(2004)\citenamefont {Larson},
  \citenamefont {Mazin},\ and\ \citenamefont {Singh}}]{larson2004magnetism}%
  \BibitemOpen
  \bibfield  {author} {\bibinfo {author} {\bibfnamefont {P.}~\bibnamefont
  {Larson}}, \bibinfo {author} {\bibfnamefont {I.}~\bibnamefont {Mazin}}, \
  and\ \bibinfo {author} {\bibfnamefont {D.}~\bibnamefont {Singh}},\
  }\href@noop {}{Magnetism, critical fluctuations, and
  susceptibility renormalization in Pd}, {\bibfield  {journal} {\bibinfo  {journal} {Phys. Rev.
  B}\ }\textbf {\bibinfo {volume} {69}},\ \bibinfo {pages} {064429} (\bibinfo
  {year} {2004})}\BibitemShut {NoStop}%
\bibitem [{\citenamefont {Sieberer}\ \emph {et~al.}(2006)\citenamefont
  {Sieberer}, \citenamefont {Khmelevskyi},\ and\ \citenamefont
  {Mohn}}]{sieberer2006magnetic}%
  \BibitemOpen
  \bibfield  {author} {\bibinfo {author} {\bibfnamefont {M.}~\bibnamefont
  {Sieberer}}, \bibinfo {author} {\bibfnamefont {S.}~\bibnamefont
  {Khmelevskyi}}, \ and\ \bibinfo {author} {\bibfnamefont {P.}~\bibnamefont
  {Mohn}},\ }\href@noop {}{Magnetic
  instability within the series TCu$_3$N (T = Pd, Rh, and Ru): A
  first-principles study}, {\bibfield  {journal} {\bibinfo  {journal} {Phys. Rev.
  B}\ }\textbf {\bibinfo {volume} {74}},\ \bibinfo {pages} {014416}
  (\bibinfo {year} {2006})}\BibitemShut {NoStop}%
\bibitem [{\citenamefont {Buschow}(1980)}]{buschow1980magnetic}%
  \BibitemOpen
  \bibfield  {author} {\bibinfo {author} {\bibfnamefont {K.}~\bibnamefont
  {Buschow}},\ }\href@noop {}{Magnetic
  properties of CeCo$_3$, Ce$_2$Co$_7$ and CeNi$_3$ and their ternary hydrides}, {\bibfield  {journal} {\bibinfo  {journal} {J.
  Less Common Met.}\ }\textbf {\bibinfo {volume} {72}},\ \bibinfo {pages} {257}
  (\bibinfo {year} {1980})}\BibitemShut {NoStop}%
\bibitem [{\citenamefont {Lemaire}(1966)}]{lemaire1966magnetic}%
  \BibitemOpen
  \bibfield  {author} {\bibinfo {author} {\bibfnamefont {R.}~\bibnamefont
  {Lemaire}},\ }\href@noop {}{Magnetic
  properties of the intermetallic compounds of cobalt with the rare earth
  metals and yttrium}, {\bibfield  {journal} {\bibinfo  {journal}
  {Cobalt}\ \textbf {\bibinfo {volume} {33}}, \ \bibinfo {pages} {201}} (\bibinfo {year} {1966})}\BibitemShut
  {NoStop}%
\bibitem [{mm-()}]{mm-note}%
  \BibitemOpen
  \href@noop {} {}\bibinfo {note} {The spin moment at Ce and Co site prsented
  in Table I is the spin moment only within the Ce or Co sphere, and does not
  include the contribution from interstitial moment. The contribution from
  interstitial and orbial moment is included in the total magnetic
  moment.}\BibitemShut {Stop}%
\bibitem [{\citenamefont {Stoner}\ \emph {et~al.}(1939)\citenamefont {Stoner}
  \emph {et~al.}}]{stoner1939collective}%
  \BibitemOpen
  \bibfield  {author} {\bibinfo {author} {\bibfnamefont {E.~C.}\ \bibnamefont
  {Stoner}} \emph {et~al.},\ }\href@noop {} {Collective electron ferromagnetism II. Energy and specific heat}, {\bibfield  {journal} {\bibinfo
  {journal} {Proc. R. Soc. Lond. A}\ }\textbf {\bibinfo {volume} {169}},\
  \bibinfo {pages} {339} (\bibinfo {year} {1939})}\BibitemShut {NoStop}%
\bibitem [{\citenamefont {Ortenzi}\ \emph {et~al.}(2011)\citenamefont
  {Ortenzi}, \citenamefont {Biermann}, \citenamefont {Andersen}, \citenamefont
  {Mazin},\ and\ \citenamefont {Boeri}}]{ortenzi2011competition}%
  \BibitemOpen
  \bibfield  {author} {\bibinfo {author} {\bibfnamefont {L.}~\bibnamefont
  {Ortenzi}}, \bibinfo {author} {\bibfnamefont {S.}~\bibnamefont {Biermann}},
  \bibinfo {author} {\bibfnamefont {O.~K.}\ \bibnamefont {Andersen}}, \bibinfo
  {author} {\bibfnamefont {I.}~\bibnamefont {Mazin}}, \ and\ \bibinfo {author}
  {\bibfnamefont {L.}~\bibnamefont {Boeri}},\ }\href@noop {}{Competition between
  electron-phonon coupling and spin fluctuations in superconducting hole-doped
  CuBiSO}, {\bibfield
  {journal} {\bibinfo  {journal} {Phys. Rev. B}\ }\textbf {\bibinfo
  {volume} {83}},\ \bibinfo {pages} {100505} (\bibinfo {year}
  {2011})}\BibitemShut {NoStop}%
\bibitem [{\citenamefont {Janak}(1977)}]{janak1977uniform}%
  \BibitemOpen
  \bibfield  {author} {\bibinfo {author} {\bibfnamefont {J.}~\bibnamefont
  {Janak}},\ }\href@noop {}{Uniform
  susceptibilities of metallic elements}, {\bibfield  {journal} {\bibinfo  {journal} {Phys.
  Rev. B}\ }\textbf {\bibinfo {volume} {16}},\ \bibinfo {pages} {255} (\bibinfo
  {year} {1977})}\BibitemShut {NoStop}%
\bibitem [{\citenamefont {Sales}(1974)}]{SalesPhD}%
  \BibitemOpen
  \bibfield  {author} {\bibinfo {author} {\bibfnamefont {B.~C.}\ \bibnamefont
  {Sales}},\ }\emph {\bibinfo {title} {Valance fluctuations on rare earth
  ions}},\ \href@noop {} {\bibinfo {type} {{PhD} dissertation}},\ \bibinfo
  {school} {University of California, San Diego} (\bibinfo {year}
  {1974})\BibitemShut {NoStop}%
\bibitem [{\citenamefont {Coey}\ \emph {et~al.}(1993)\citenamefont {Coey},
  \citenamefont {Allan}, \citenamefont {Minakov},\ and\ \citenamefont
  {Bugaslavsky}}]{coey1993ce2fe17}%
  \BibitemOpen
  \bibfield  {author} {\bibinfo {author} {\bibfnamefont {J.}~\bibnamefont
  {Coey}}, \bibinfo {author} {\bibfnamefont {J.}~\bibnamefont {Allan}},
  \bibinfo {author} {\bibfnamefont {A.}~\bibnamefont {Minakov}}, \ and\
  \bibinfo {author} {\bibfnamefont {Y.~V.}\ \bibnamefont {Bugaslavsky}},\
  }\href@noop {}{Ce$_2$Fe$_{17}$: Mixed valence or 4f band?}, {\bibfield  {journal} {\bibinfo  {journal} {J. Appl. Phys.}\ }\textbf {\bibinfo {volume} {73}},\ \bibinfo {pages} {5430}
  (\bibinfo {year} {1993})}\BibitemShut {NoStop}%
\bibitem [{\citenamefont {Jarlborg}(2014)}]{jarlborg2014role}%
  \BibitemOpen
  \bibfield  {author} {\bibinfo {author} {\bibfnamefont {T.}~\bibnamefont
  {Jarlborg}},\ }\href@noop {}{Role of
  thermal disorder for magnetism and the $\alpha$-$\gamma$ transition in
  cerium: Results from density-functional theory}, {\bibfield  {journal} {\bibinfo  {journal}
  {Phys. Rev. B}\ }\textbf {\bibinfo {volume} {89}},\ \bibinfo {pages} {184426}
  (\bibinfo {year} {2014})}\BibitemShut {NoStop}%
\bibitem [{\citenamefont {Alam}\ and\ \citenamefont
  {Johnson}(2014)}]{alam2014mixed}%
  \BibitemOpen
  \bibfield  {author} {\bibinfo {author} {\bibfnamefont {A.}~\bibnamefont
  {Alam}}\ and\ \bibinfo {author} {\bibfnamefont {D.~D.}\ \bibnamefont
  {Johnson}},\ }\href@noop {}{Mixed valency
  and site-preference chemistry for cerium and its compounds: A predictive
  density-functional theory study}, {\bibfield  {journal} {\bibinfo  {journal} {Phys.
  Rev. B}\ }\textbf {\bibinfo {volume} {89}},\ \bibinfo {pages} {235126}
  (\bibinfo {year} {2014})}\BibitemShut {NoStop}%
\bibitem [{\citenamefont {Matsumoto}\ \emph {et~al.}(2009)\citenamefont
  {Matsumoto}, \citenamefont {Han}, \citenamefont {Otsuki},\ and\ \citenamefont
  {Savrasov}}]{matsumoto2009first}%
  \BibitemOpen
  \bibfield  {author} {\bibinfo {author} {\bibfnamefont {M.}~\bibnamefont
  {Matsumoto}}, \bibinfo {author} {\bibfnamefont {M.~J.}\ \bibnamefont {Han}},
  \bibinfo {author} {\bibfnamefont {J.}~\bibnamefont {Otsuki}}, \ and\ \bibinfo
  {author} {\bibfnamefont {S.~Y.}\ \bibnamefont {Savrasov}},\ }\href@noop {} {First-principles simulations of heavy fermion cerium compounds based on the
  kondo lattice},
  {\bibfield  {journal} {\bibinfo  {journal} {Phys. Rev. Lett.}\ }\textbf
  {\bibinfo {volume} {103}},\ \bibinfo {pages} {096403} (\bibinfo {year}
  {2009})}\BibitemShut {NoStop}%
\bibitem [{\citenamefont {Sales}\ and\ \citenamefont
  {Viswanathan}(1976)}]{sales1976demagnetization}%
  \BibitemOpen
  \bibfield  {author} {\bibinfo {author} {\bibfnamefont {B.}~\bibnamefont
  {Sales}}\ and\ \bibinfo {author} {\bibfnamefont {R.}~\bibnamefont
  {Viswanathan}},\ }\href@noop {} {Demagnetization due to interconfiguration fluctuations in the RE-Cu$_2$Si$_2$ compounds}, {\bibfield  {journal} {\bibinfo  {journal}
  {J. Low Temp. Phys.}\ }\textbf {\bibinfo {volume} {23}},\ \bibinfo {pages}
  {449} (\bibinfo {year} {1976})}\BibitemShut {NoStop}%
\bibitem [{\citenamefont {Sales}\ and\ \citenamefont
  {Wohlleben}(1975)}]{sales1975susceptibility}%
  \BibitemOpen
  \bibfield  {author} {\bibinfo {author} {\bibfnamefont {B.}~\bibnamefont
  {Sales}}\ and\ \bibinfo {author} {\bibfnamefont {D.}~\bibnamefont
  {Wohlleben}},\ }\href@noop {}{Susceptibility of interconfiguration-fluctuation compounds}, {\bibfield  {journal} {\bibinfo  {journal}
  {Phys. Rev. Lett.}\ }\textbf {\bibinfo {volume} {35}},\ \bibinfo {pages}
  {1240} (\bibinfo {year} {1975})}\BibitemShut {NoStop}%
\bibitem [{\citenamefont {Sales}(1977)}]{sales1977model}%
  \BibitemOpen
  \bibfield  {author} {\bibinfo {author} {\bibfnamefont {B.~C.}\ \bibnamefont
  {Sales}},\ }\href@noop {}{A model for the
  thermodynamic properties of metallic rare earth systems with an unstable
  valence}, {\bibfield  {journal} {\bibinfo  {journal} {J. Low
  Temp. Phys.}\ }\textbf {\bibinfo {volume} {28}},\ \bibinfo {pages} {107}
  (\bibinfo {year} {1977})}\BibitemShut {NoStop}%
\bibitem [{\citenamefont {Pandey}\ \emph {et~al.}(2018)\citenamefont {Pandey},
  \citenamefont {Du},\ and\ \citenamefont {Parker}}]{pandey_2_17_3_paper}%
  \BibitemOpen
  \bibfield  {author} {\bibinfo {author} {\bibfnamefont {T.}~\bibnamefont
  {Pandey}}, \bibinfo {author} {\bibfnamefont {M.-H.}\ \bibnamefont {Du}}, \
  and\ \bibinfo {author} {\bibfnamefont {D.~S.}\ \bibnamefont {Parker}},\
  }\href@noop {} {Tuning the magnetic properties and structural stabilities of the 2-17-3
  magnets Sm$_2$Fe$_{17}$X$_3$ (X = C, N) by substituting La or Ce for Sm}, {\bibfield  {journal} {\bibinfo  {journal} {Phys. Rev. Appl.}\
  }\textbf {\bibinfo {volume} {9}},\ \bibinfo {pages} {034002} (\bibinfo {year}
  {2018})}\BibitemShut {NoStop}%
\bibitem [{\citenamefont {Pandey}\ and\ \citenamefont
  {Parker}(2018)}]{pandey2018magnetic}%
  \BibitemOpen
  \bibfield  {author} {\bibinfo {author} {\bibfnamefont {T.}~\bibnamefont
  {Pandey}}\ and\ \bibinfo {author} {\bibfnamefont {D.~S.}\ \bibnamefont
  {Parker}},\ }\href@noop {}{Magnetic properties and magnetocrystalline anisotropy of Nd$_2$Fe$_{17}$, Nd$_2$Fe$_{17}$X$_3$, and related compounds}, {\bibfield  {journal} {\bibinfo  {journal} {Sci.
  Rep.}\ }\textbf {\bibinfo {volume} {8}},\ \bibinfo {pages} {3601} (\bibinfo
  {year} {2018})}\BibitemShut {NoStop}%
\bibitem [{\citenamefont {Larson}\ and\ \citenamefont
  {Mazin}(2003{\natexlab{b}})}]{larson2003magnetic}%
  \BibitemOpen
  \bibfield  {author} {\bibinfo {author} {\bibfnamefont {P.}~\bibnamefont
  {Larson}}\ and\ \bibinfo {author} {\bibfnamefont {I.}~\bibnamefont {Mazin}},\
  }\href@noop {}{Magnetic properties of
  SmCo$_5$ and YCo$_5$}, {\bibfield  {journal} {\bibinfo  {journal} {J. Appl. Phys.}\
  }\textbf {\bibinfo {volume} {93}},\ \bibinfo {pages} {6888} (\bibinfo {year}
  {2003}{\natexlab{b}})}\BibitemShut {NoStop}%
\bibitem [{\citenamefont {Zhu}\ \emph {et~al.}(2014)\citenamefont {Zhu},
  \citenamefont {Janoschek}, \citenamefont {Rosenberg}, \citenamefont
  {Ronning}, \citenamefont {Thompson}, \citenamefont {Torrez}, \citenamefont
  {Bauer},\ and\ \citenamefont {Batista}}]{zhu2014lda}%
  \BibitemOpen
  \bibfield  {author} {\bibinfo {author} {\bibfnamefont {J.-X.}\ \bibnamefont
  {Zhu}}, \bibinfo {author} {\bibfnamefont {M.}~\bibnamefont {Janoschek}},
  \bibinfo {author} {\bibfnamefont {R.}~\bibnamefont {Rosenberg}}, \bibinfo
  {author} {\bibfnamefont {F.}~\bibnamefont {Ronning}}, \bibinfo {author}
  {\bibfnamefont {J.}~\bibnamefont {Thompson}}, \bibinfo {author}
  {\bibfnamefont {M.~A.}\ \bibnamefont {Torrez}}, \bibinfo {author}
  {\bibfnamefont {E.~D.}\ \bibnamefont {Bauer}}, \ and\ \bibinfo {author}
  {\bibfnamefont {C.~D.}\ \bibnamefont {Batista}},\ }\href@noop {}{LDA+DMFT approach to
  magnetocrystalline anisotropy of strong magnets}, {\bibfield
  {journal} {\bibinfo  {journal} {Phys. Rev. X}\ }\textbf {\bibinfo {volume}
  {4}},\ \bibinfo {pages} {021027} (\bibinfo {year} {2014})}\BibitemShut
  {NoStop}%
\bibitem [{\citenamefont {Daalderop}\ \emph {et~al.}(1994)\citenamefont
  {Daalderop}, \citenamefont {Kelly},\ and\ \citenamefont
  {Schuurmans}}]{daalderop1994magnetic}%
  \BibitemOpen
  \bibfield  {author} {\bibinfo {author} {\bibfnamefont {G.}~\bibnamefont
  {Daalderop}}, \bibinfo {author} {\bibfnamefont {P.}~\bibnamefont {Kelly}}, \
  and\ \bibinfo {author} {\bibfnamefont {M.}~\bibnamefont {Schuurmans}},\
  }\href@noop {}{Magnetic anisotropy of a
  free-standing Co monolayer and of multilayers which contain Co monolayers}, {\bibfield  {journal} {\bibinfo  {journal} {Phys. Rev.
  B}\ }\textbf {\bibinfo {volume} {50}},\ \bibinfo {pages} {9989} (\bibinfo
  {year} {1994})}\BibitemShut {NoStop}%
\bibitem [{\citenamefont {Antropov}\ \emph {et~al.}(2014)\citenamefont
  {Antropov}, \citenamefont {Ke},\ and\ \citenamefont
  {{\AA}berg}}]{antropov2014constituents}%
  \BibitemOpen
  \bibfield  {author} {\bibinfo {author} {\bibfnamefont {V.}~\bibnamefont
  {Antropov}}, \bibinfo {author} {\bibfnamefont {L.}~\bibnamefont {Ke}}, \ and\
  \bibinfo {author} {\bibfnamefont {D.}~\bibnamefont {{\AA}berg}},\ }\href@noop
  {}{Constituents of magnetic anisotropy and a screening of spin-orbit coupling
  in solids}, {\bibfield  {journal} {\bibinfo  {journal} {Solid State Commun.}\
  }\textbf {\bibinfo {volume} {194}},\ \bibinfo {pages} {35} (\bibinfo {year}
  {2014})}\BibitemShut {NoStop}%
\bibitem [{\citenamefont {Ke}\ and\ \citenamefont {van
  Schilfgaarde}(2015)}]{ke2015band}%
  \BibitemOpen
  \bibfield  {author} {\bibinfo {author} {\bibfnamefont {L.}~\bibnamefont
  {Ke}}\ and\ \bibinfo {author} {\bibfnamefont {M.}~\bibnamefont {van
  Schilfgaarde}},\ }\href@noop {}{Band-filling effect on magnetic anisotropy using a Green$^{\prime}$s function
  method}, {\bibfield  {journal} {\bibinfo  {journal}
  {Phys. Rev. B}\ }\textbf {\bibinfo {volume} {92}},\ \bibinfo {pages}
  {014423} (\bibinfo {year} {2015})}\BibitemShut {NoStop}%
\bibitem [{\citenamefont {Ke}\ \emph {et~al.}(2016)\citenamefont {Ke},
  \citenamefont {Kukusta},\ and\ \citenamefont {Johnson}}]{ke2016origin}%
  \BibitemOpen
  \bibfield  {author} {\bibinfo {author} {\bibfnamefont {L.}~\bibnamefont
  {Ke}}, \bibinfo {author} {\bibfnamefont {D.}~\bibnamefont {Kukusta}}, \ and\
  \bibinfo {author} {\bibfnamefont {D.~D.}\ \bibnamefont {Johnson}},\
  }\href@noop {}{Origin of magnetic
  anisotropy in doped Ce$_2$Co$_{17}$ alloys}, {\bibfield  {journal} {\bibinfo  {journal} {Phys. Rev. B}\ }\textbf {\bibinfo {volume} {94}},\ \bibinfo {pages} {144429} (\bibinfo
  {year} {2016})}\BibitemShut {NoStop}%
\bibitem [{\citenamefont {Bruno}(1989)}]{bruno1989tight}%
  \BibitemOpen
  \bibfield  {author} {\bibinfo {author} {\bibfnamefont {P.}~\bibnamefont
  {Bruno}},\ }\href@noop {}{Tight-binding
  approach to the orbital magnetic moment and magnetocrystalline anisotropy of
  transition-metal monolayers}, {\bibfield  {journal} {\bibinfo  {journal}
  {Phys. Rev. B}\ }\textbf {\bibinfo {volume} {39}},\ \bibinfo {pages}
  {865} (\bibinfo {year} {1989})}\BibitemShut {NoStop}%
\bibitem [{\citenamefont {Skubic}\ \emph {et~al.}(2008)\citenamefont {Skubic},
  \citenamefont {Hellsvik}, \citenamefont {Nordstr{\"o}m},\ and\ \citenamefont
  {Eriksson}}]{skubic2008method}%
  \BibitemOpen
  \bibfield  {author} {\bibinfo {author} {\bibfnamefont {B.}~\bibnamefont
  {Skubic}}, \bibinfo {author} {\bibfnamefont {J.}~\bibnamefont {Hellsvik}},
  \bibinfo {author} {\bibfnamefont {L.}~\bibnamefont {Nordstr{\"o}m}}, \ and\
  \bibinfo {author} {\bibfnamefont {O.}~\bibnamefont {Eriksson}},\ }\href@noop
  {} {A method for atomistic spin dynamics simulations: implementation and
  examples}, {\bibfield  {journal} {\bibinfo  {journal} {J. Phys. Condens. Matter}\ }\textbf {\bibinfo {volume} {20}},\ \bibinfo {pages} {315203}
  (\bibinfo {year} {2008})}\BibitemShut {NoStop}%
\bibitem [{\citenamefont {Evans}\ \emph {et~al.}(2014)\citenamefont {Evans},
  \citenamefont {Fan}, \citenamefont {Chureemart}, \citenamefont {Ostler},
  \citenamefont {Ellis},\ and\ \citenamefont {Chantrell}}]{evans2014atomistic}%
  \BibitemOpen
  \bibfield  {author} {\bibinfo {author} {\bibfnamefont {R.~F.}\ \bibnamefont
  {Evans}}, \bibinfo {author} {\bibfnamefont {W.~J.}\ \bibnamefont {Fan}},
  \bibinfo {author} {\bibfnamefont {P.}~\bibnamefont {Chureemart}}, \bibinfo
  {author} {\bibfnamefont {T.~A.}\ \bibnamefont {Ostler}}, \bibinfo {author}
  {\bibfnamefont {M.~O.}\ \bibnamefont {Ellis}}, \ and\ \bibinfo {author}
  {\bibfnamefont {R.~W.}\ \bibnamefont {Chantrell}},\ }\href@noop {}{Atomistic spin model simulations of magnetic nanomaterials}, {\bibfield
   {journal} {\bibinfo  {journal} {J. Phys. Condens. Matter}\
  }\textbf {\bibinfo {volume} {26}},\ \bibinfo {pages} {103202} (\bibinfo
  {year} {2014})}\BibitemShut {NoStop}%
\bibitem [{\citenamefont {Eriksson}\ \emph {et~al.}(2017)\citenamefont
  {Eriksson}, \citenamefont {Bergman}, \citenamefont {Bergqvist},\ and\
  \citenamefont {Hellsvik}}]{eriksson2017atomistic}%
  \BibitemOpen
  \bibfield  {author} {\bibinfo {author} {\bibfnamefont {O.}~\bibnamefont
  {Eriksson}}, \bibinfo {author} {\bibfnamefont {A.}~\bibnamefont {Bergman}},
  \bibinfo {author} {\bibfnamefont {L.}~\bibnamefont {Bergqvist}}, \ and\
  \bibinfo {author} {\bibfnamefont {J.}~\bibnamefont {Hellsvik}},\ }\href@noop
  {} {\emph {\bibinfo {title} {Atomistic spin dynamics: foundations and
  applications}}}\ (\bibinfo  {publisher} {Oxford university press},\ \bibinfo
  {year} {2017})\BibitemShut {NoStop}%
\bibitem [{\citenamefont {Jernberg}\ \emph {et~al.}(1984)\citenamefont
  {Jernberg}, \citenamefont {Yousif}, \citenamefont {H{\"a}ggstr{\"o}m},\ and\
  \citenamefont {Andersson}}]{jernberg1984}%
  \BibitemOpen
  \bibfield  {author} {\bibinfo {author} {\bibfnamefont {P.}~\bibnamefont
  {Jernberg}}, \bibinfo {author} {\bibfnamefont {A.}~\bibnamefont {Yousif}},
  \bibinfo {author} {\bibfnamefont {L.}~\bibnamefont {H{\"a}ggstr{\"o}m}}, \
  and\ \bibinfo {author} {\bibfnamefont {Y.}~\bibnamefont {Andersson}},\
  }\href@noop {}{A m{\"o}ssbauer study of Fe$_2$P$_{1-x}$Si$_x$ (x $\leqslant$ 0.35)}, {\bibfield  {journal} {\bibinfo  {journal} {J. Solid State
  Chem.}\ }\textbf {\bibinfo {volume} {53}},\ \bibinfo {pages} {313} (\bibinfo
  {year} {1984})}\BibitemShut {NoStop}%
\bibitem [{\citenamefont {Delczeg-Czirjak}\ \emph {et~al.}(2010)\citenamefont
  {Delczeg-Czirjak}, \citenamefont {Delczeg}, \citenamefont {Punkkinen},
  \citenamefont {Johansson}, \citenamefont {Eriksson},\ and\ \citenamefont
  {Vitos}}]{delczeg2010}%
  \BibitemOpen
  \bibfield  {author} {\bibinfo {author} {\bibfnamefont {E.~K.}\ \bibnamefont
  {Delczeg-Czirjak}}, \bibinfo {author} {\bibfnamefont {L.}~\bibnamefont
  {Delczeg}}, \bibinfo {author} {\bibfnamefont {M.~P.~J.}\ \bibnamefont
  {Punkkinen}}, \bibinfo {author} {\bibfnamefont {B.}~\bibnamefont
  {Johansson}}, \bibinfo {author} {\bibfnamefont {O.}~\bibnamefont {Eriksson}},
  \ and\ \bibinfo {author} {\bibfnamefont {L.}~\bibnamefont {Vitos}},\
  }\href@noop {}{Ab initio study of structural and magnetic properties of Si-doped Fe$_2$P}, {\bibfield  {journal} {\bibinfo  {journal} {Phys. Rev. B}\
  }\textbf {\bibinfo {volume} {82}},\ \bibinfo {pages} {085103} (\bibinfo
  {year} {2010})}\BibitemShut {NoStop}%
\bibitem [{\citenamefont {Pandey}\ and\ \citenamefont
  {Parker}()}]{pandey2018b}%
  \BibitemOpen
  \bibfield  {author} {\bibinfo {author} {\bibfnamefont {T.}~\bibnamefont
  {Pandey}}\ and\ \bibinfo {author} {\bibfnamefont {D.~S.}\ \bibnamefont
  {Parker}},\ }\href@noop {}{Magnteic properties of Fe$_2$Ta under hafnium doping}, {\bibinfo  {journal} {In preparation}\
  }\BibitemShut {NoStop}%
\bibitem [{\citenamefont {Coey}(2012)}]{coey2012}%
  \BibitemOpen
\bibfield  {journal} {  }\bibfield  {author} {\bibinfo {author} {\bibfnamefont
  {J.}~\bibnamefont {Coey}},\ }\href@noop {}{Permanent magnets: Plugging the gap}, {\bibfield  {journal} {\bibinfo
  {journal} {Scr. Mater.}\ }\textbf {\bibinfo {volume} {67}},\ \bibinfo {pages}
  {524} (\bibinfo {year} {2012})}\BibitemShut {NoStop}%
\bibitem [{\citenamefont {K{\"u}bler}(2017)}]{kubler2017theory}%
  \BibitemOpen
  \bibfield  {author} {\bibinfo {author} {\bibfnamefont {J.}~\bibnamefont
  {K{\"u}bler}},\ }\href@noop {} {\emph {\bibinfo {title} {Theory of itinerant
  electron magnetism}}},\ Vol.\ \bibinfo {volume} {106}\ (\bibinfo  {publisher}
  {Oxford University Press},\ \bibinfo {year} {2017})\BibitemShut {NoStop}%
\bibitem [{\citenamefont {Zhuang}\ \emph {et~al.}(2016)\citenamefont {Zhuang},
  \citenamefont {Kent},\ and\ \citenamefont {Hennig}}]{zhuang2016strong}%
  \BibitemOpen
  \bibfield  {author} {\bibinfo {author} {\bibfnamefont {H.~L.}\ \bibnamefont
  {Zhuang}}, \bibinfo {author} {\bibfnamefont {P.}~\bibnamefont {Kent}}, \ and\
  \bibinfo {author} {\bibfnamefont {R.~G.}\ \bibnamefont {Hennig}},\
  }\href@noop {} {Strong anisotropy and magnetostriction in the two-dimensional stoner ferromagnet Fe$_3$GeTe$_2$}, {\bibfield  {journal} {\bibinfo  {journal} {Phys. Rev.
  B}\ }\textbf {\bibinfo {volume} {93}},\ \bibinfo {pages} {134407} (\bibinfo
  {year} {2016})}\BibitemShut {NoStop}%
\bibitem [{\citenamefont {Jin}\ \emph {et~al.}(2016)\citenamefont {Jin},
  \citenamefont {Zhang}, \citenamefont {Bai}, \citenamefont {Qian},
  \citenamefont {Wu}, \citenamefont {Ma}, \citenamefont {Shen},\ and\
  \citenamefont {Yan}}]{jin2016manipulating}%
  \BibitemOpen
  \bibfield  {author} {\bibinfo {author} {\bibfnamefont {J.}~\bibnamefont
  {Jin}}, \bibinfo {author} {\bibfnamefont {Y.}~\bibnamefont {Zhang}}, \bibinfo
  {author} {\bibfnamefont {G.}~\bibnamefont {Bai}}, \bibinfo {author}
  {\bibfnamefont {Z.}~\bibnamefont {Qian}}, \bibinfo {author} {\bibfnamefont
  {C.}~\bibnamefont {Wu}}, \bibinfo {author} {\bibfnamefont {T.}~\bibnamefont
  {Ma}}, \bibinfo {author} {\bibfnamefont {B.}~\bibnamefont {Shen}}, \ and\
  \bibinfo {author} {\bibfnamefont {M.}~\bibnamefont {Yan}},\ }\href@noop {}{Manipulating Ce valence in RE$_2$Fe$_{14}$B tetragonal compounds by La-Ce co-doping: Resultant crystallographic and magnetic anomaly},
  {\bibfield  {journal} {\bibinfo  {journal} {Sci. Rep.}\ }\textbf
  {\bibinfo {volume} {6}},\ \bibinfo {pages} {30194} (\bibinfo {year}
  {2016})}\BibitemShut {NoStop}%
\bibitem [{\citenamefont {Alam}\ \emph {et~al.}(2013)\citenamefont {Alam},
  \citenamefont {Khan}, \citenamefont {McCallum},\ and\ \citenamefont
  {Johnson}}]{alam2013site}%
  \BibitemOpen
  \bibfield  {author} {\bibinfo {author} {\bibfnamefont {A.}~\bibnamefont
  {Alam}}, \bibinfo {author} {\bibfnamefont {M.}~\bibnamefont {Khan}}, \bibinfo
  {author} {\bibfnamefont {R.~W.}\ \bibnamefont {McCallum}}, \ and\ \bibinfo
  {author} {\bibfnamefont {D.~D.}\ \bibnamefont {Johnson}},\ }\href@noop {}{Site-preference and valency for rare-earth sites in (R-Ce)$_2$Fe$_{14}$B magnets},
  {\bibfield  {journal} {\bibinfo  {journal} {Appl. Phys. Lett.}\
  }\textbf {\bibinfo {volume} {102}},\ \bibinfo {pages} {042402} (\bibinfo
  {year} {2013})}\BibitemShut {NoStop}%
\bibitem [{\citenamefont {Isnard}\ \emph {et~al.}(1998)\citenamefont {Isnard},
  \citenamefont {Miraglia}, \citenamefont {Guillot},\ and\ \citenamefont
  {Fruchart}}]{isnard1998hydrogen}%
  \BibitemOpen
  \bibfield  {author} {\bibinfo {author} {\bibfnamefont {O.}~\bibnamefont
  {Isnard}}, \bibinfo {author} {\bibfnamefont {S.}~\bibnamefont {Miraglia}},
  \bibinfo {author} {\bibfnamefont {M.}~\bibnamefont {Guillot}}, \ and\
  \bibinfo {author} {\bibfnamefont {D.}~\bibnamefont {Fruchart}},\ }\href@noop
  {}{Hydrogen effects on the magnetic properties of RFe$_{11}$Ti compounds}, {\bibfield  {journal} {\bibinfo  {journal} {J. Alloys Compd.}\ }\textbf {\bibinfo {volume} {275}},\ \bibinfo {pages} {637}
  (\bibinfo {year} {1998})}\BibitemShut {NoStop}%
\bibitem [{\citenamefont {Chaboy}\ \emph {et~al.}(1995)\citenamefont {Chaboy},
  \citenamefont {Marcelli}, \citenamefont {Bozukov}, \citenamefont {Baudelet},
  \citenamefont {Dartyge}, \citenamefont {Fontaine},\ and\ \citenamefont
  {Pizzini}}]{chaboy1995effect}%
  \BibitemOpen
  \bibfield  {author} {\bibinfo {author} {\bibfnamefont {J.}~\bibnamefont
  {Chaboy}}, \bibinfo {author} {\bibfnamefont {A.}~\bibnamefont {Marcelli}},
  \bibinfo {author} {\bibfnamefont {L.}~\bibnamefont {Bozukov}}, \bibinfo
  {author} {\bibfnamefont {F.}~\bibnamefont {Baudelet}}, \bibinfo {author}
  {\bibfnamefont {E.}~\bibnamefont {Dartyge}}, \bibinfo {author} {\bibfnamefont
  {A.}~\bibnamefont {Fontaine}}, \ and\ \bibinfo {author} {\bibfnamefont
  {S.}~\bibnamefont {Pizzini}},\ }\href@noop {}  {Effect of hydrogen absorption on the cerium electronic state in CeFe$_{11}$Ti:
  An x-ray-absorption and circular-magnetic-dichroism investigation}, {\bibfield  {journal} {\bibinfo
   {journal} {Phys. Rev. B}\ }\textbf {\bibinfo {volume} {51}},\ \bibinfo
  {pages} {9005} (\bibinfo {year} {1995})}\BibitemShut {NoStop}%
\end{thebibliography}
%

\end{document}